%% file: paper.tex
\documentclass[letterpaper]{sig-alternate-10pt}
\usepackage[pass]{geometry}

\usepackage{epsfig}
\usepackage[outdir=./]{epstopdf}

\usepackage{url}
\usepackage{microtype}
\usepackage{epsfig}
\usepackage{color}
\usepackage{pseudocode}
\usepackage{listings}
\usepackage{multirow}
\usepackage{setspace}
\usepackage{csquotes}

\usepackage{caption}
\usepackage{subcaption}

\usepackage{blindtext}
\usepackage{etoolbox}
\usepackage{xifthen}
\usepackage{setspace}

\makeatletter
\patchcmd{\maketitle}{}{}{}
\makeatother

\begin{document}

\title{Evolving Twitter: an experimental analysis of graph properties of the social graph.}
 
\numberofauthors{3}

\author{
\alignauthor Despoina Antonakaki\titlenote{Despoina Antonakaki is also with the University of Crete.}\\
       \affaddr{FORTH-ICS, Greece}\\
       \email{despoina@ics.forth.gr}
\and
\alignauthor Sotiris Ioannidis\\
       \affaddr{FORTH-ICS, Greece}\\
       \email{sotiris@ics.forth.gr}
\and
\alignauthor Paraskevi Fragopoulou\\
       \affaddr{FORTH-ICS, Greece}\\
       \email{fragopou@ics.forth.gr}
}

\maketitle

\begin{abstract}

Twitter is one of the most prominent Online Social Networks. It covers a significant part of the online worldwide 
population(\textasciitilde20\%) and has impressive growth rates. The social graph of Twitter has been the subject of numerous studies since it can reveal the intrinsic properties of large and complex online communities. Despite the plethora of these studies, there is a limited cover on the properties of the social graph while they evolve over time. Moreover, due to the extreme size of this social network (millions of nodes, billions of edges), there is a small subset of possible graph properties that can be efficiently measured in a reasonable timescale.
In this paper we propose a sampling framework that allows the estimation of graph properties on large social networks. We apply this framework to a subset of Twitter's social network that has 13.2 million users, 8.3 billion edges and covers the complete Twitter timeline (from April 2006 to January 2015). We derive estimation on the time evolution of 24 graph properties many of which have never been measured on large social networks. We further discuss how these estimations shed more light on the inner structure and growth dynamics of Twitter's social network.
\end{abstract}

\input{introduction}
\input{acknowledgements}

\bibliographystyle{acm}
\bibliography{paper}


\end{document}

%% file: introduction.tex
\section{Introduction}
\label{sec:introduction}
Twitter is a popular microblogging social platform, established on 2006 and as of today has reached 645 million registered users \cite{twitterStatistics} where half of them are monthly active. 
Except from ordinary individual users, Twitter is utilized from news agents, public figures, and organizations to disseminate their activity and engage in discussion with other users. 
The online activity and the dynamics of the social graph of Twitter are considered to be indicative of the tendencies of the off-line social life and reflect the preferences of the public in general \cite{Java2007}.
For these reasons the structure and properties of the social graph of Twitter has been the subject of numerous studies that seek to model and sometimes predict the behaviour of users as well as how this behaviour affects the growth dynamics of the graph. 

An online social network (OSN) represents users as nodes and user relations as edges.
The most famous and well established feature of OSNs are the scale-free structure \cite{Ferrara2012,milgram1967small,travers1969experimental,newman2006structure,albert1999internet}.
Alternative this feature is coined as `small-world' structure and is associated with the six-degrees of separation attribute.
This feature is attributed to other features of OSNs as for example the lifetime of a tweet through re-tweets \cite{retweet1}.

A model that describes accurately the structure of an OSN can be of extreme importance. 
For example \cite{barbieri2014follow} suggested a following recommendation system based on social information on common graph properties as well as a community detection method \cite{barbieri2013cascade}. 
Other metrics that measure the popularity and the impact of a user's activity (for example betweenness centrality) can be of extreme importance for evaluation of marketing, political or personal campaigns. 
Policy makers can utilize these metrics in order to increase their influence~\cite{2_bray_2013}. 

One of the most important consideration when performing these studies is the extreme computational requirements.
Since the algorithms for extracting these properties can hardly be parallelized the graph has to be loaded in memory.
Thus, an OSN that has a scale-free structure can exceed the memory of an advanced computer (i.e. 64GB of RAM) with a very small proportion of the complete graph (i.e. 10 million nodes).
It is indicative that the number of active Twitter users every month reaches approximately 302 million. 
Another consideration is the high computational complexity of some metrics. 
Although some metrics require time linear to the number of nodes, other essential metrics require quadratic or higher time (Table~\ref{tab:allmetrics}).


In this paper we present an empirical analysis of the evolution of 24 graph metrics on the social graph of Twitter.
For this purpose we have taken a sample of Twitter's OSN with 13.2 million users that contain approximately 8.3 billion following relationships. 
All relationships are sorted according to an estimation of the link creation time. 
Twitter's API does not provide the following creation time.
Nevertheless we applied a heuristic that computes a lower bound of this creation time~\cite{meeder2011we}.
This heuristic is based on the fact that the lists of followers is returned sorted according to creation time from Twitter's API.
With this information we can approximate the OSN of Twitter while it evolved from Twitter's beginning (April 2006) until today.

We also present a sampling framework on Twitter's OSN suitable for estimation of graph metrics.
This framework is based on the fact that although some metrics are practically impossible to be computed on large network, a random sub-sampling of the network can, most of the times, give a good approximation.
Moreover some metrics accept a `cutoff' parameter (i.e. betweenness) that eases the computation and returns an approximation of the metric.
We present the values of all these 24 metrics and their evolution in our dataset through time. 

Based on these measurements we are able to identify three crucial time periods with different growth dynamics. 
These periods suggest that there was an inflationary, a deflationary and a still going on stable growth rate on Twitter.











\subsection{Major findings and Organization}
\label{sec:organization}
The major findings of this paper are the following:
 \begin{itemize}
	 \item We apply a heuristic that allows the estimation of the link creation of the following relations on Twitter's OSN. This allows the split of our dataset in various datapoints (or eras) and the measurement of the graph metrics to each one of them. 
	 \item We apply a massive graph analysis by applying graph measures from all the spectrum of available metrics many of which have not been studied before on Twitter's OSN.
	 \item We present a two-dimensional sampling method. The first dimension is the time and the second are subsets of Nodes or subgraphs depending on the time requirements of the graph metrics.
	 \item Through these measurements we assess the structural evolution of Twitter's graph along with the evolution of user specific metrics.  
	 \end{itemize}

The rest of this paper is organized as follows: On section~\ref{sec:introduction} we present a background of existing studies on graph measurement and sampling of large OSNs.
On section~\ref{sec:datacollection} we present out data collection methods and the basic nature of the collected data. We also perform an initial analysis of followbacks and the size of the largest components within our collected data.
On section~\ref{sec:properties} we proceed to the analysis of all different graph metrics. Subsection~\ref{sec:measurementmethods} presents our sampling technique and other heuristics that allowed the assessment of metrics with extreme time requirements.
Finally on section~\ref{sec:discussion} we discuss our findings, we present some limitation of the current study along with our priorities on future work.   

\section{Background}
\label{sec:introduction}

Graph metrics on OSNs provide valuable insights on their structure, evolution and modelling.
In \cite{myers2014information} the authors have studied the degree distribution, connected components, shortest path lengths, clustering coefficient, two-hop neighborhood and assortativity metrics in a subset of 175 million active Twitter users. 
One of the most significant conclusions from this study regarding the structure of Twitter (as well as other social networks) is that the distribution of the network's nodes degree follows a power law. 
According to graph theory this means that the network has a scale-free structure \cite{milgram1967small,travers1969experimental}.
Other principles of OSNs are the six degrees of separation \cite{newman2006structure,albert1999internet} and the strength of weak ties \cite{granovetter1973strength,szell2010multirelational}.
The scale-free structure governs not only the nodes degrees but also Twitter's reply network \cite{bliss2014evolutionary,Kwak2010}.
Regarding modelling, \cite{PhysRevX.4.031046} studied the evolution of average shortest path and average degree to model the underlying social structure of a network that governs its evolution. 
These metrics have been used before \cite{leskovec2007graph} to generate graph generation models that resemble OSNs and other `real-life' graphs. 
Metrics like density, clustering, heterogeneity and modularity have been used to study the evolution of OSNs~\cite{Hu20091105}. 

Graph metrics can also give insights regarding a user's activity and popularity in a network \cite{morales2014efficiency}. 
Since the number of followers has been valued as an insufficient measure of a user's popularity \cite{cha2010measuring,10.1371/journal.pone.0084265} other metrics are taken into use.
For example \cite{1_englehardt_2015} studies the ``Betweenness Centrality" metric which measures the level in which a user is in the center of her local network.
Others metrics that measures a user's popularity is `pagerank' \cite{Kwak2010} and `centrality' \cite{abs-1209-4616}. 

Another area where graph metrics can give insights of an OSNs is security.
\cite{wang2014whispers} have studied the degree distribution, clustering coefficient, average path length and assortativity on the anonymous social network `Whispers' and identify, besides its evolution dynamics, some vulnerabilities that can expose user's identity. Other areas are community detection \cite{Fortunato2010} and follow recommendation systems \cite{barbieri2014follow}.

It is essential to note that we can extract valuable information about Twitter's OSN without using sophisticated graph metrics.
In a very well designed study, \cite{Kwak2010} crawled that entire Twitter site as of July 2009.
They measured the user's friends, followers and tweets distribution as well as the reciprocity level (ratio of follow-back relations) and homophily (the rate at which similar people interact compared to dissimilar people~\cite{2062393}).
They also came to useful conclusions regarding the social impact of Twitter by measuring the trends and retweets distributions. 
From graph metrics, they measured the degree distribution and average shortest path.

\subsection{Sampling large OSNs}
\label{sec:sampling}
As we have discussed, sampling is a necessary step in order to apply even simple metrics on extremely large OSNs \cite{Lovasz2009}.
To our knowledge, the best review on sampling techniques and their efficacy is \cite{Leskovec2006}.
The most important finding of this study is that a 15\% sample size is usually enough for the estimation of most of the real graph properties.
They assess various sampling techniques from which the best are `Random Walk' and `Forest Fires'. 
Random Walk is when we select a random node and then we simulate a random walk. At each step there is a probability (c=0.15) of returning to the starting node and repeating the procedure.
Forest Fire simulates a `fire'. Starting from a random node we `burn' a subset of its edges. We proceed recursively to burn part of the adjacent edges of the edges that are already burned. By controlling the forward and backward burning probabilities we can generate a subset of a given size~\cite{Leskovec2005}. 
Although they assess the efficacy of these sampling techniques on many networks, none of their experiments includes an online social network.
Moreover the simple Random Nodes technique that we apply in 6 out of the total 24 metrics (4 by choosing randomly single nodes and 2 by making random subgraphs), although it is not the best, has an efficiency that is close to the best chosen (0.272 for Random Nodes, compared to 0.202 of Random Walk).
Random Nodes seem to exhibit the highest bias on the representation of the distribution of sizes of weakly connected components.
A set of nodes are weekly connected if there exists an undirected path from any pair of nodes in the set.
In another study \cite{gabielkov2014sampling} the authors try two different approaches: obtaining most popular users and obtaining an unbiased sample of users.
They argue that the best unbiased sampling technique is to query the social network for randomly generated IDs, or else Random Nodes. 
Nevertheless we plan to apply more sophisticated sampling techniques in our future work.









\section{Data Collection}
\label{sec:datacollection}
Our data consist of a list of Twitter's users, their followers and their followings. 
We used Twitter's API to collect data for our experiments. 
We started by collecting the followers and following list of the Twitter account of the corresponding author of the present paper (@antonakd).
We continued with a recursive approach, namely we collected the followers and followings of the users in the existing followers and followings lists.
Our biggest impediment in this process was the limitations of the Twitter API.
Twitter allows 15 queries per 15 minutes from a single account. Moreover each query can return maximum the IDs of 5.000 users. 
To put this throttle in perspective, one account requires approximately one year to get the followers and followings of 17.500 users assuming that none of them has more than 5.000 friends or followers.
If a user has 50 million followers (like many celebrities and organization) it takes a week to get the complete follower list of this user. 
To overcome this we set up various Twitter accounts and in total we generated 1.250 downloading applications in a period of two months trying to be very considerate on Twitter's terms of service. 
Moreover we didn't collect the followers of users with more than 5.000 followers and we marked these users as ``celebrities".
Beside the overall speedup of our data collection process, this exclusion results in a social graph without nodes with extreme degrees.
According to a Cumulative Distribution Function (CDF) of the distribution of the number of followers, the percentage of users with more than 5.000 followers is less than 1\% \cite{Stringhini2013}.  
In total we downloaded the friends and followers list of 13.2 million users of which 154.318 were marked as celebrities (1.1\%).
The resulting social graph has 8.3 billion edges. 
Before Applying an evolutionary study on the social graph of Twitter, it is necessary the ordering of the edges according to creation time. 

Although Twitter does not reveal the creation time of followings we can apply a heuristic that produces a lower bound estimation \cite{meeder2011we}.
This heuristic is based on the fact that Twitter's API returns the lists of friends and followers of a user ordered by creation time. 
We also know that user's IDs are ordered according to account creation time. 
If we apply to this knowledge the simple intuition that a following to or from an account happens after this account is created we can infer the following heuristic:
If users $U_1$, $U_2$, ..,$U_n$ followed user A in that order then a lower bound estimation of the $U_n$ $\rightarrow$ A following relationship, is the most recent creation time of the accounts $U_1, U_2, .., U_n$. 
Or else, an estimation of the creation time of a following is the most recent account creation time of the users that also did a following prior to this. 
\cite{meeder2011we} proved that this heuristic is pretty accurate specially on time periods where there are high follow rates. 
We applied this heuristic on our dataset and we ordered all 8.3 billion following relationships according to this. 

\subsection{Followbacks}
\label{sec:followbacks}
\begin{figure}[t]
  \includegraphics[width=0.45\textwidth]{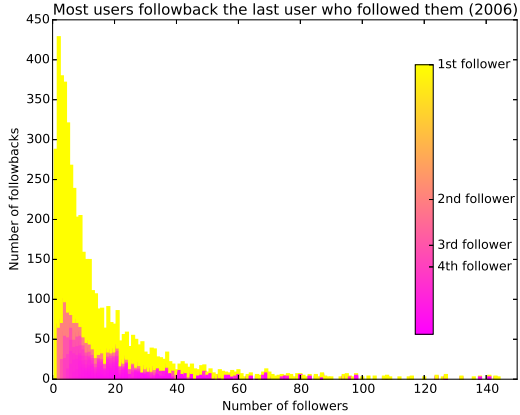}
	\caption{Most users followback the last user that followed them (yellow area)}
	\label{fig:followbacks}
\end{figure}

Our first experiment was the investigation of the order of the follow-backs that occurred in our dataset.
The main reason for this was to study the accuracy of the following time creation heuristic. 
Let's assume that we have two events: The first is that A follows B and the second is that user B follows back user A. 
The question is how many other users did user B followed in the time period between the two events.
If there are no users that means user B followed A right after user A followed her (1st follower). If there is one user, then she follow-backed the 2nd follower who followed her, etc.  
On figure~\ref{fig:followbacks} we plot this order. 
We applied this analysis only in 2006 because the followings are more sparse and the heuristic is considered more inaccurate. 
Nevertheless by doing this, we confirm our intuition that the vast majority of follow-backs happened to the very first (most recent) user that followed us.
This finding can be applied to a recommendation system as a generic guideline: `Suggest the users that recently followed a user'.

\subsection{Size of largest connected graph}
\label{sec:connectedgraph}
\begin{figure}[t]
  \includegraphics[width=0.45\textwidth]{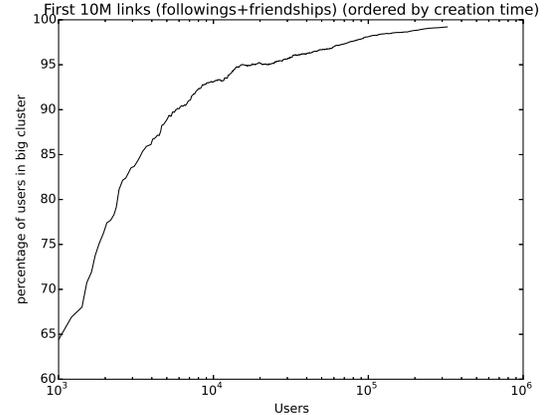}
	\caption{CDF of the ratio of users that belong to the largest connected graph.}
	\label{fig:connected}
\end{figure}

Our second experiment was to determine the size of largest subgraph component.
The question here is: When does the OSN of Twitter evolves to a point that it becomes a connected graph?
Starting from Twitter's creation time (April 2006) we measured the ratio of nodes that belonged to the largest connected subgraph to the complete number of nodes.
A graph is connected when there exist a path from any node to any other node in the graph. 
For this study we ignore directions and treat each edge as undirected.
On figure~\ref{fig:connected} we plot this measure for the first 10 million followings, or else from April 2006 to July 2008. 
We notice that even at the beginning of Twitter and given our subset of Twitter's OSN, more than 95\% of users belong to the largest connected graph.
Practically we can assume that, in our dataset, after 2007 the OSN of Twitter is a connected graph. 

\section{Graph Properties}
\label{sec:properties}
\begin{figure}[t]
  \includegraphics[width=0.45\textwidth]{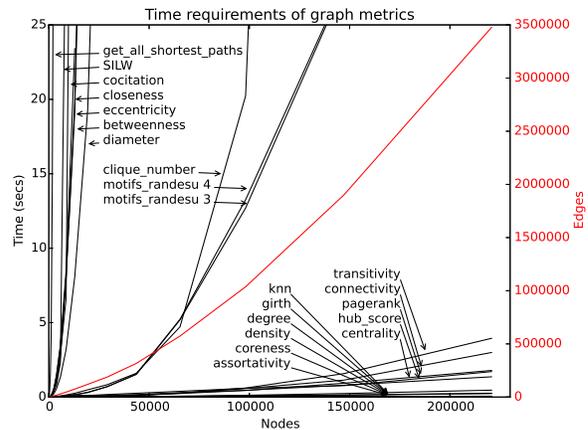}
	\caption{The time requirements for all graph metrics. In this graph we exclude `neighborhood size' and `strength' which require minimal computational time. The red line shows the number of edges of the graph}
	\label{fig:requirements}
\end{figure}

For our analysis we chose 24 different graph properties.
These metrics are implemented in igraph~\cite{igraph} which is a high performance graph library that has also bindings in the python language. 
Initially me measured the time requirements for each of the metrics.
On figure~\ref{fig:requirements} we plot these measurements for a progressively growing social graph of Twitter (starting from April 2006) until it reaches 250.000 nodes. 
We notice that we have a family of metrics that have exponential time requirements. 
These metrics are all\_shortest\_paths, similarity inverse log weighted (SILW), co-citation, closeness, eccentricity, betweenness and diameter.
There is a family that follows time complexity proportional to the number of edges like clique number and motifs RAND-ESU.
Finally a big family of metrics requires minimal time proportional to the time of nodes or lower. 
This family includes degree, density, coreness, assortativity, transitivity, connectivity, pagerank, hub\_score, centrality, strength and neighborhood size. 

\subsection{Measurement Methods}
\label{sec:measurementmethods}
Having an approximation of the link creation time in our dataset, allows us to perform an evolution study of all graph metrics.
Namely, we can measure these metrics and study their evolution while the graph grows over time. 
To tackle the extreme computational requirements we split the graph in various time points. 
We initially investigate the early phase of Twitter by measuring these metrics for each month from April 2006 to December 2008.
After that we split the graph per day from 1st of January 2009 to 1st of January 2015. 
Although our dataset contains connections timed after January 2015 we removed these to eliminate possible batch effects in our analysis. 

For metrics that do not require excessive time we simply measure their values.
Metrics that require excessive computation time we apply one of the following techniques:
In cases where a metric can be computed in a subset of nodes we randomly choose 1000 nodes and we measure the metric.
Then we measure the 95\% confidence interval of this sampling. 
This gives an indication of how precise our estimation was. 
If the confidence interval is too large (greater than half of the mean value of the measurement) then we repeat this procedure until we get a confidence interval that is shorter than half of the mean value of the measurement.
If the total time spent on an estimation is more than 2 hours then we stop and report the existing confidence intervals.
In cases that a metric cannot be applied to certain node but requires a complete graph, we create 1000 random subplots of size 100 and we apply the same confidence interval estimation as described above. 
Then we continue by increasing the size of the sampled subplot by a factor of 1.5. 
That means that initially we take 1000 random subplots of size 100, then the size increases to 150, then to 225 etc. 
We stop this procedure when the size of the subplot reaches the size of the real graph or after 2 hours of computation.
Finally there are cases where the measured metrics accepts a `cutoff' parameter. 
This is the maximum size of the path that it should consider when measuring this metric.
In these cases we progressively apply cutoff values starting from 2 until again the limit of 2 hours of computation is reached. 

With these estimations we try to utilize the available computation power without relying in a-priori sampling methods.
On Table~\ref{tab:allmetrics} we present all metrics, their time complexity according to the igraph authors and the sampling technique that we applied.

\begin{table}[h]
\begin{tabular}{|l|l|l|}
\hline
\multicolumn{1}{|c|}{\textbf{Metric}} & \multicolumn{1}{c|}{\textbf{\begin{tabular}[c]{@{}c@{}}Time\\ Complexity\end{tabular}}} & \multicolumn{1}{c|}{\textbf{Sampling}} \\ \hline
Assortativity                         & O($|$E$|$)                                                                                  & None                                   \\ \hline
Betweenness                           & O($|$V$|$*$|$E$|$)                                                                              & Cutoff                                 \\ \hline
Cliques                               & O(3\textasciicircum ($|$V$|$/3))                                                            & Subgraph                               \\ \hline
Closeness                             & O(n$|$E$|$)                                                                                 & Cutoff                                 \\ \hline
Cocitation                            & O($|$V$|$d\textasciicircum 2)                                                               & rnd nodes                           \\ \hline
Coreness                              & O($|$E$|$)                                                                                  & None                                   \\ \hline
Degree dstr                           & O($|$V$|$)                                                                                  & None                                   \\ \hline
Density                               & O(1)                                                                                    & None                                   \\ \hline
Diameter                              & O($|$V$|$*$|$E$|$)                                                                              & Subgraph                               \\ \hline
Eccentricity                          & O(n*($|$V$|$+$|$E$|$))                                                                          & rnd nodes                           \\ \hline
Edge connectivity                     & O($|$V$|$\textasciicircum 4)                                                                & None                                   \\ \hline
Eigenvector centrality                & O($|$V$|$+$|$E$|$)                                                                              & None                                   \\ \hline
All shortest paths                    & O(n!)                                                                                   & rnd nodes                           \\ \hline
Hub score                             & O($|$V$|$)                                                                                  & None                                   \\ \hline
KNN                                   & O($|$V$|$+$|$E$|$)                                                                              & None                                   \\ \hline
Max degree                            & O($|$V$|$)                                                                                  & None                                   \\ \hline
Motifs rand-ESU 3                     & NA                                                                                      & None                                   \\ \hline
Motifs rand-ESU 4                     & NA                                                                                      & None                                   \\ \hline
Neighborhood size                     & O(n*d*o)                                                                                & None                                   \\ \hline
Pagerank                              & O($|$V$|$+$|$E$|$)                                                                              & None                                   \\ \hline
SILW                                  & O($|$V$|$d\textasciicircum 2)                                                               & rnd nodes                           \\ \hline
Strength                              & O($|$V$|$+$|$E$|$)                                                                              & None                                   \\ \hline
Transitivity local                          & O($|$V$|$*d\textasciicircum 2)                                                              & None                                   \\ \hline
Transitivity global                         & O($|$V$|$*d\textasciicircum 2)                                                              & None                                   \\ \hline
\end{tabular}
\caption{List of evaluated metrics. On the time complexity column, $|$V$|$ is the number of nodes, $|$E$|$ is the number of edges, d is the graph's maximum degree, n is the number of nodes for which this metric is applied and o is the order. The third column contains the sample techniques used for computational intense metrics.}
\label{tab:allmetrics}
\end{table}


\subsection{Assortativity}
\label{sec:Assortativity}
\begin{figure}[t]
  \centering
  \includegraphics[width=0.48\textwidth]{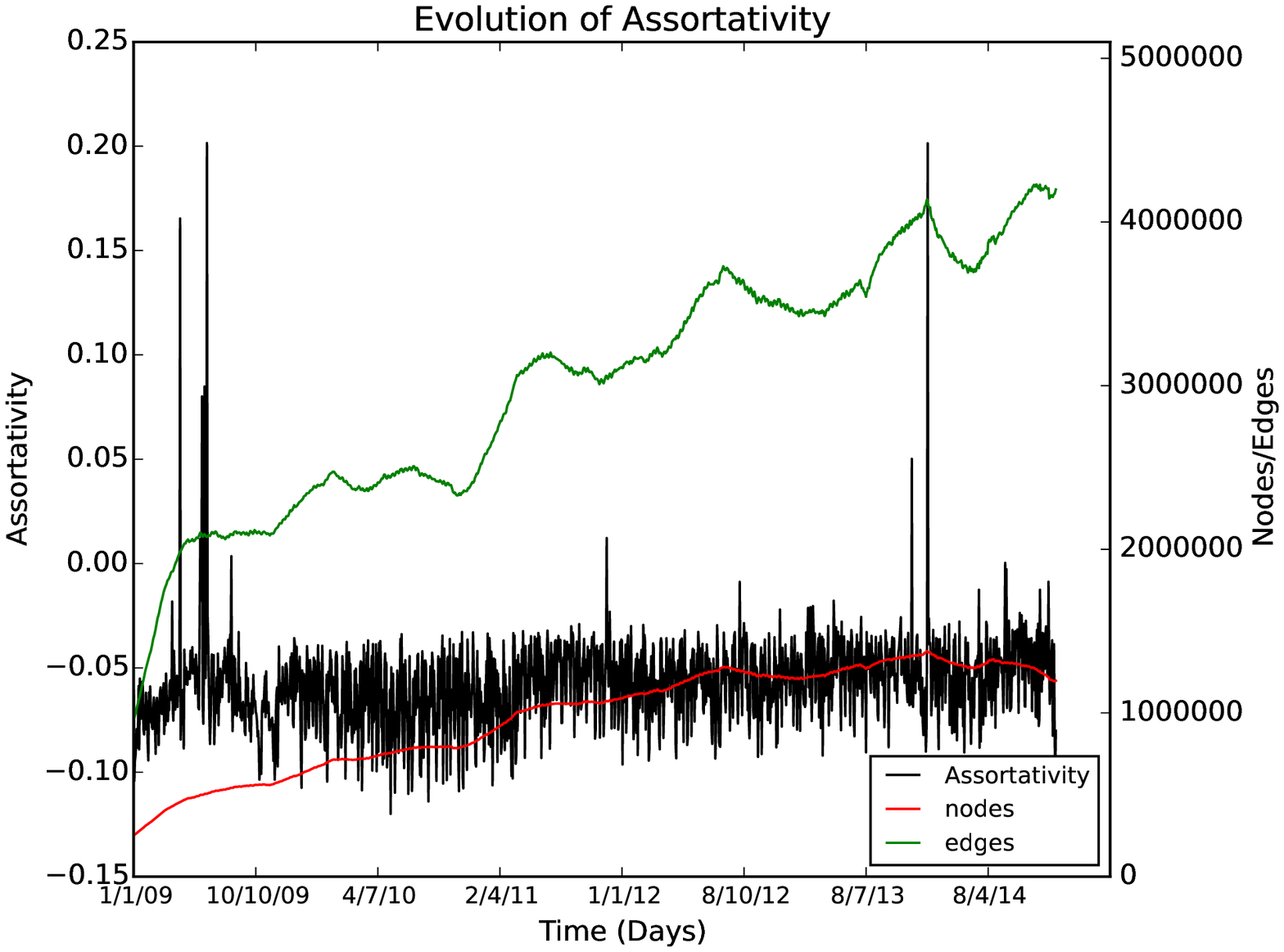}\llap{\makebox[12.3cm]{\raisebox{3.95cm}{\includegraphics[height=2cm]{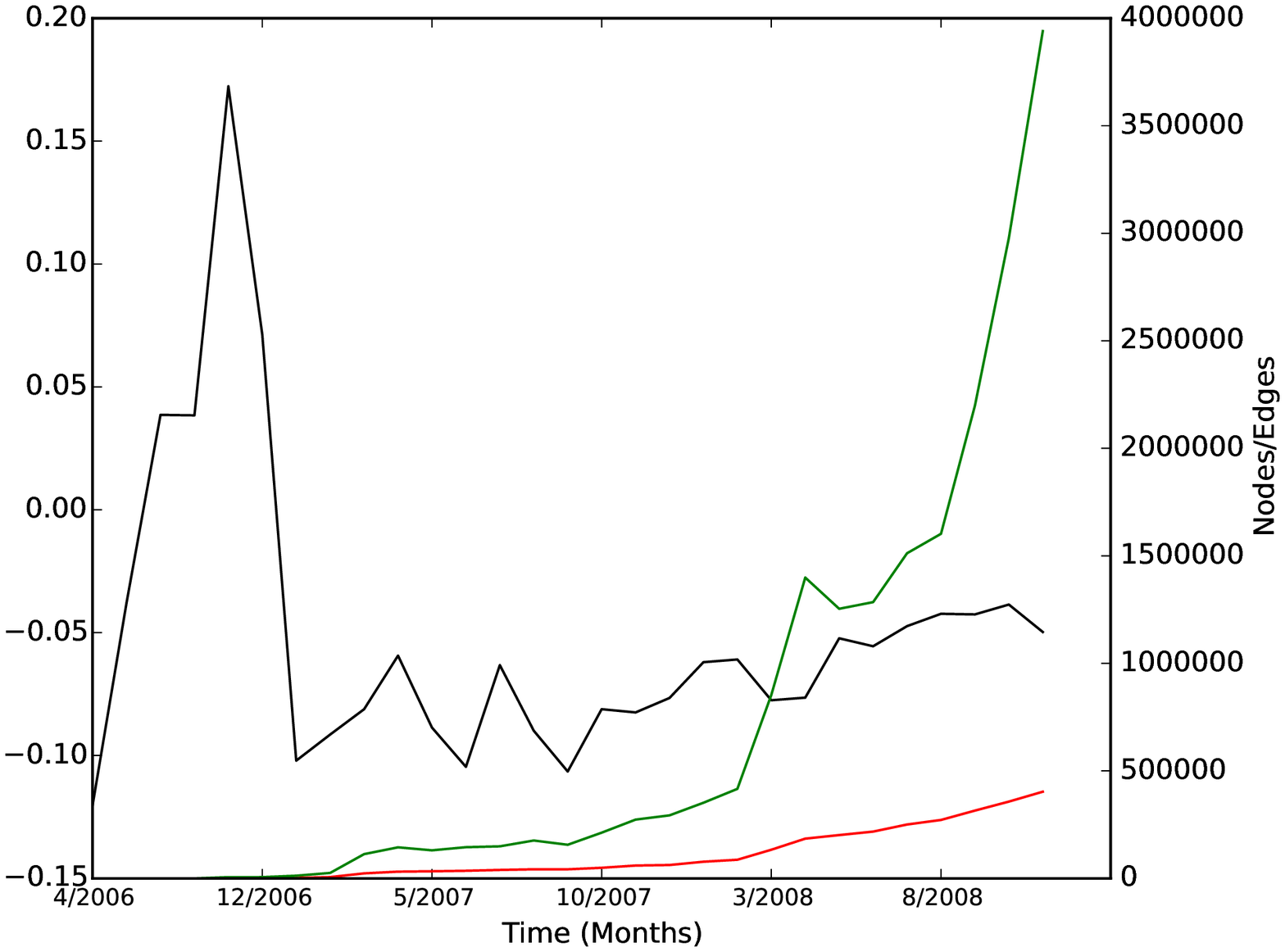}}}}
  \caption{Assortativity degree per day from start of 2009 to the end of 2014 and per month from April 2006 to December 2009 (embedded graph).}
  \label{fig:assortativity}
\end{figure}

Assortativity measures the degree of which nodes with some properties tend to connect with nodes with similar properties.
In our case the property that we measure is the nodes degree. 
Zero assortativity shows no correlation, 1 shows highest assortativity and -1 shows dis-assortativity (meaning that nodes with low degree tend to connect with nodes with high degree).
In figure~\ref{fig:assortativity} we show that in general the OSN of Twitter shows a small degree of dis-assortativity that is constant throughout time.

\subsection{Betweenness}
\label{sec:betweenness}
\begin{figure}[t]
  \centering
  \includegraphics[width=0.48\textwidth]{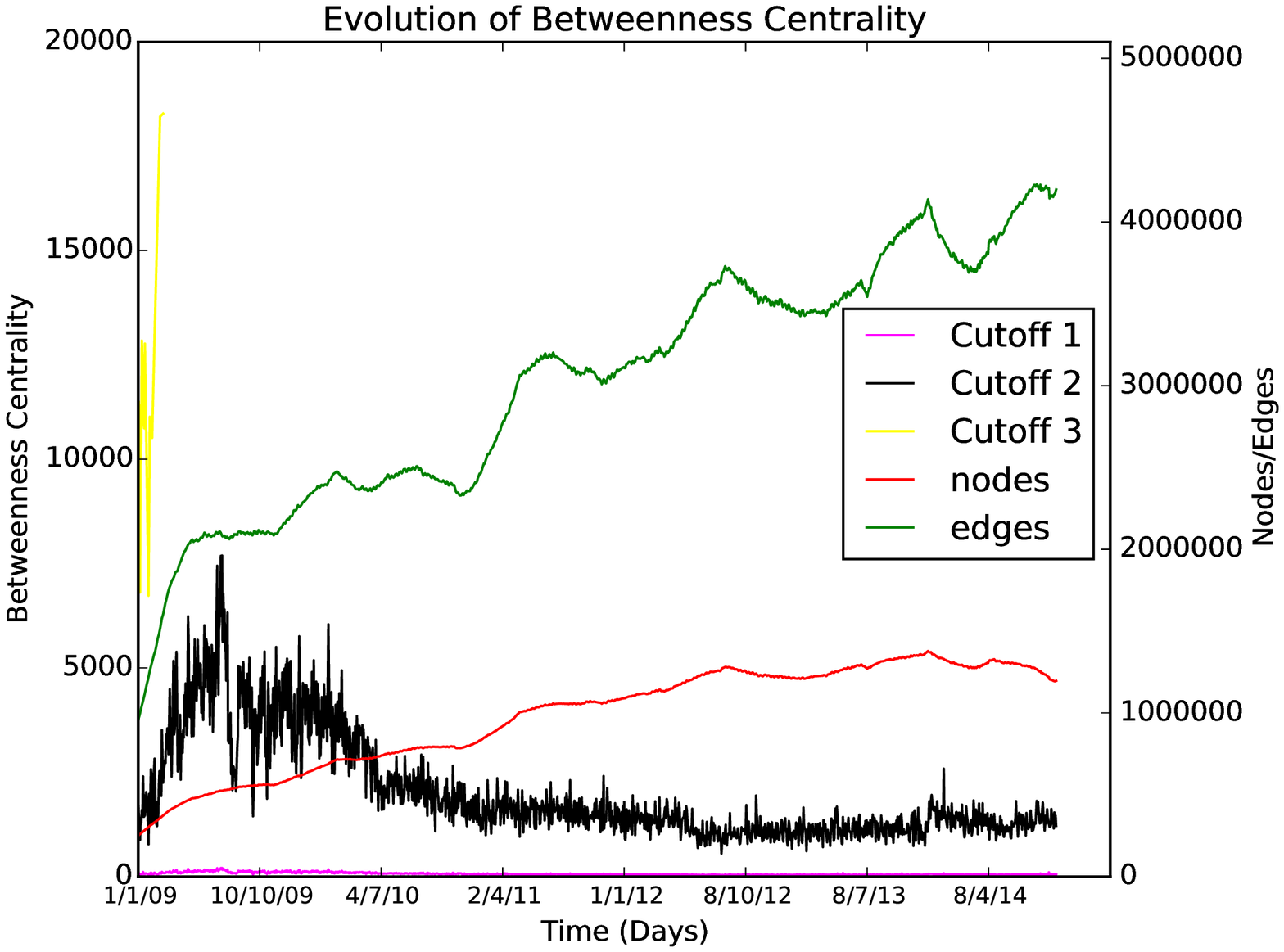}\llap{\makebox[12.0cm]{\raisebox{3.95cm}{\includegraphics[height=2cm]{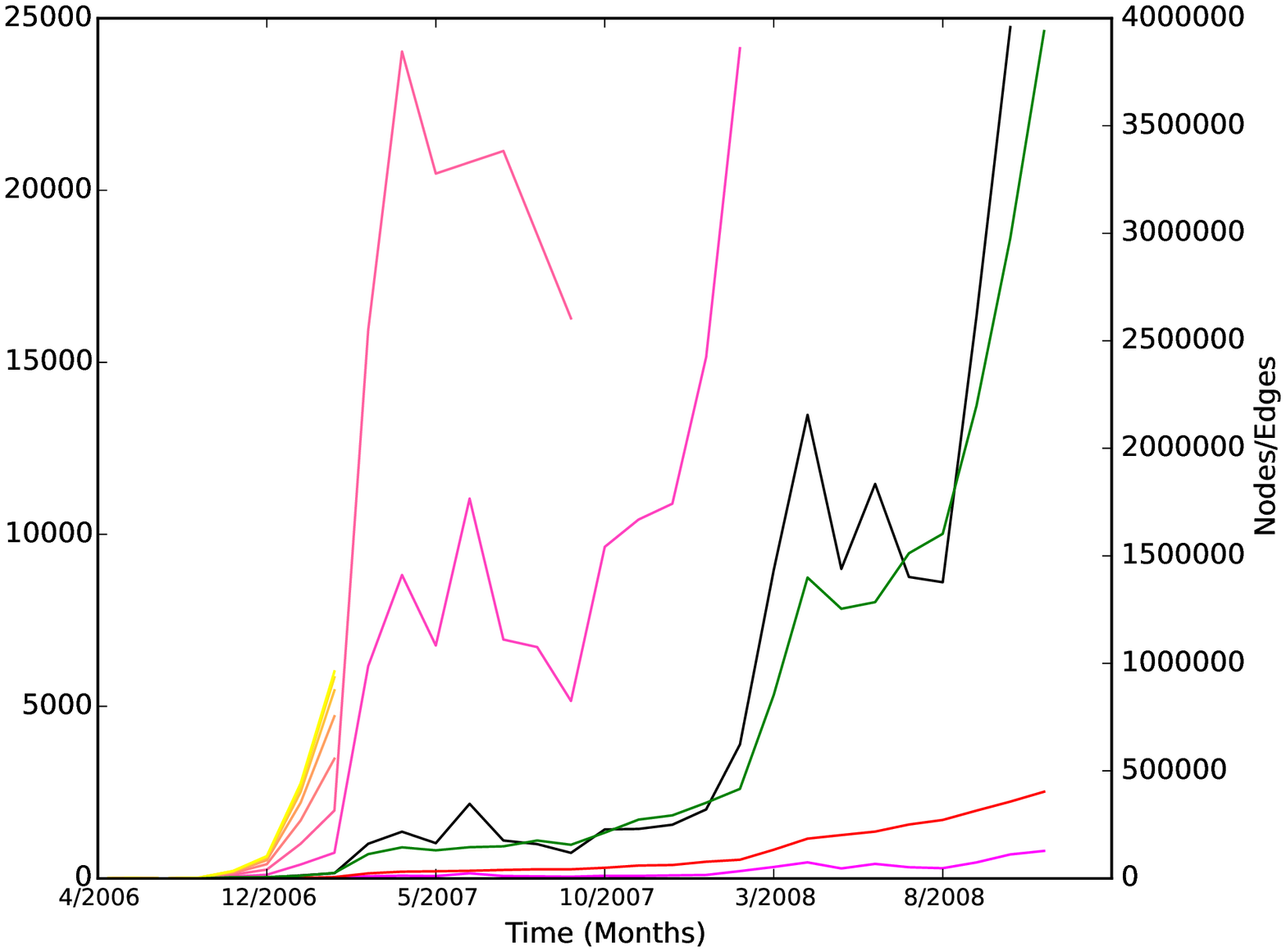}}}}
  \caption{Betweenness Centrality per day from start of 2009 to the end of 2014 and per month from April 2006 to December 2009 (embedded graph).}
  \label{fig:betweenness}
\end{figure}

Betwenness centrality is a measure of how central a node is on a graph. 
It is defined as the number of shortest paths from any node to any other node that pass from this node.
The measure of centrality is of great importance when measuring the impact of a user in her local network under the assumption that information flow follows the shortest path.
On figure ~\ref{fig:betweenness} we plot the average betweenness centrality over all nodes for each day from 2009 to the end of 2014. For computational reasons we accounted only for paths of length 2 (black line). 
We notice that the network experiences a period at the beginning where the centrality increases. 
After the end of 2009 the centrality drops and stabilizes to the number of 1000 after 2011. 
We also notice this increase in the embedded plot that shows the betwenness centrality increase for the beginning of Twitter (April 2006 to December 2009). 
The centrality follows the same rate of increase as the number of edges in the graph.

\subsection{Maximum Clique}
\label{sec:maxclique}
\begin{figure}[t]
  \centering
  \includegraphics[width=0.48\textwidth]{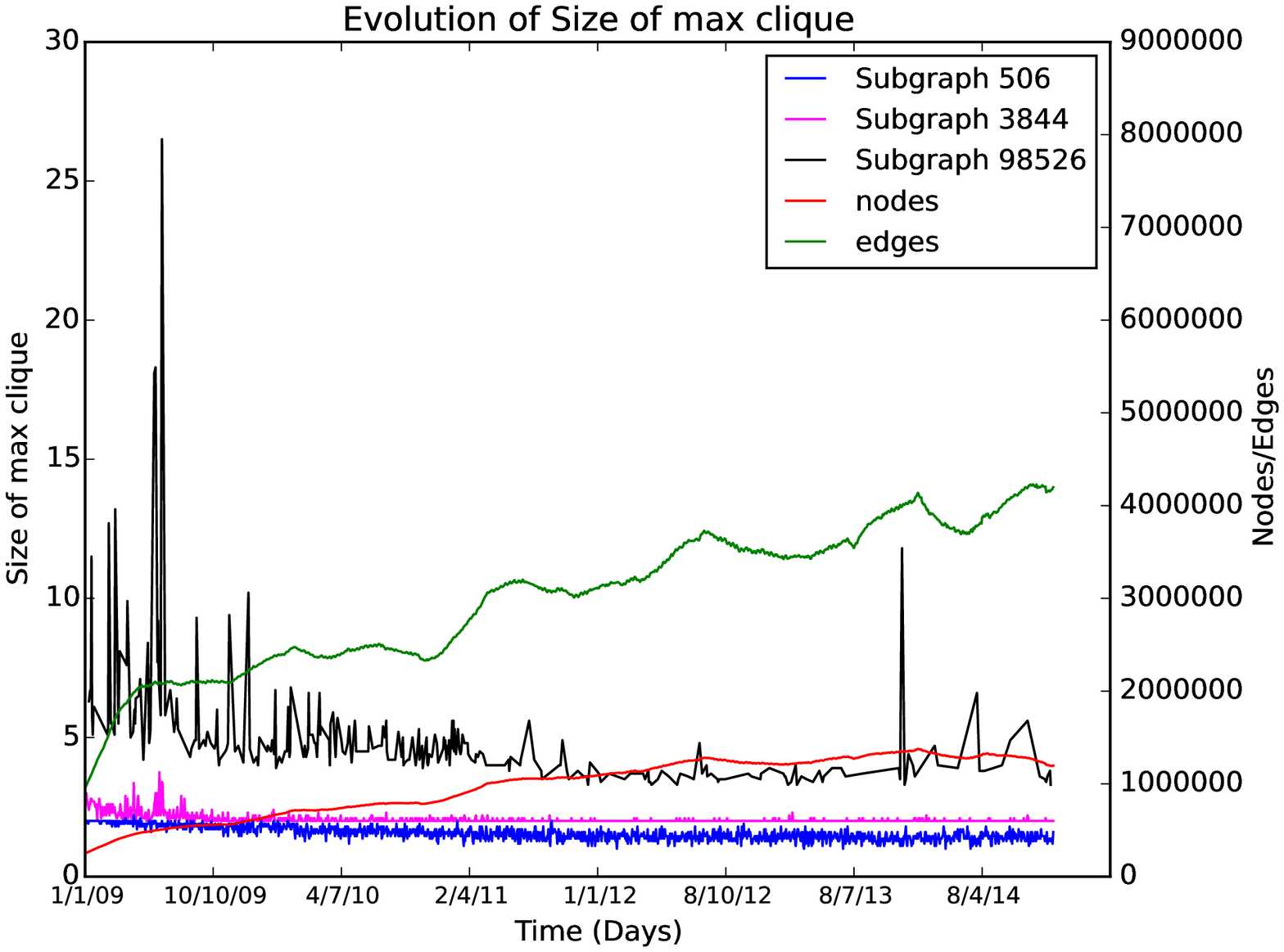}\llap{\makebox[11.62cm]{\raisebox{2.9cm}{\includegraphics[height=3cm]{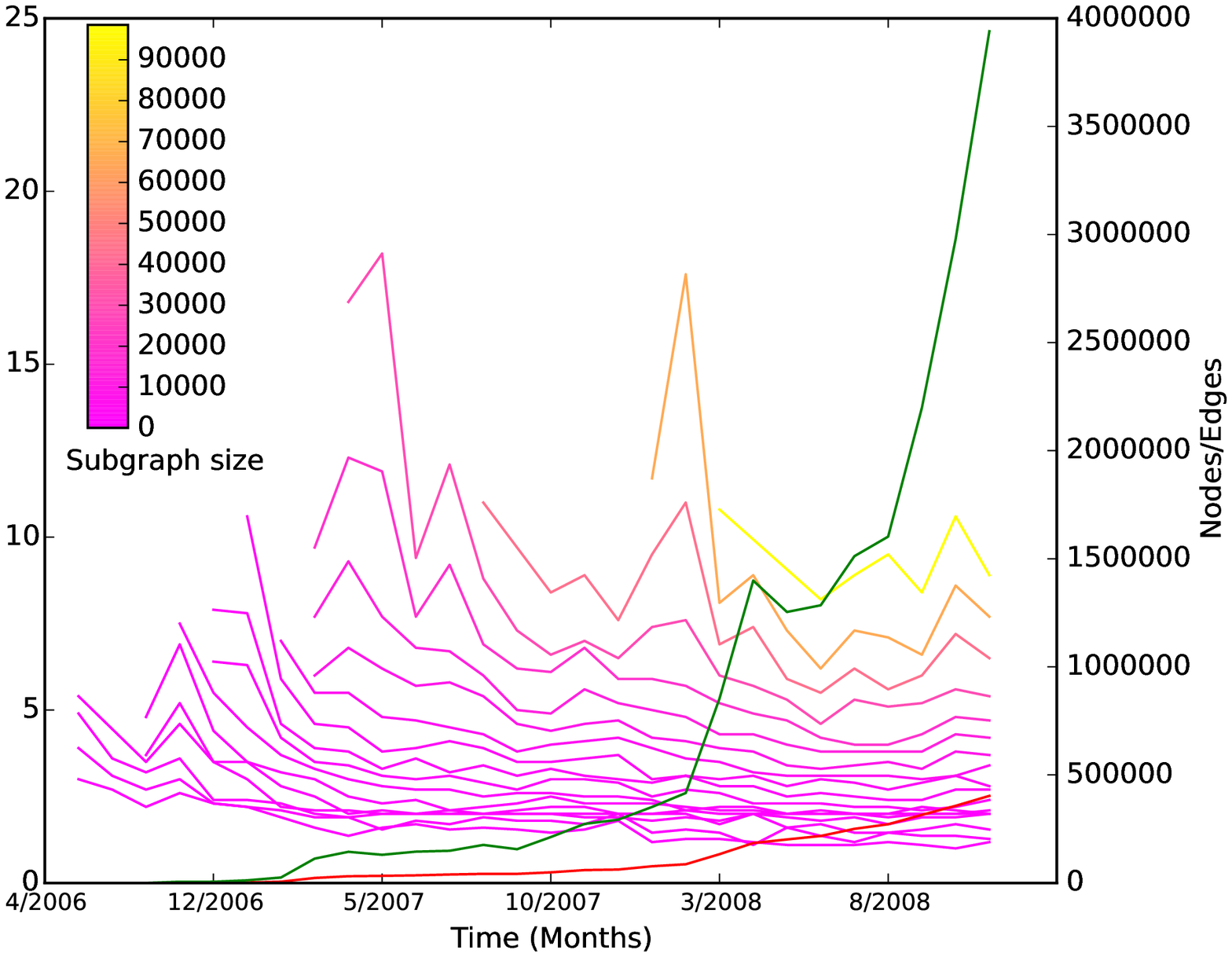}}}}
  \caption{Maximum cliques per day from start of 2009 to the end of 2014 and per month from April 2006 to December 2009 (embedded graph). The colorscale shows the number of sampled nodes in the Random Nodes sub-sampling procedure.}
  \label{fig:maxclique}
\end{figure}

A clique in a graph is a subgraph that is fully connected, or else there exist an edge for every pair of nodes in the subgraph. 
Here we search for the size of the maximum clique in the graph.
The computation cost of this process is cubic to the size of edges. 
For this reason we applied the subgraph sampling presented in section~\ref{sec:measurementmethods}.
In figure~\ref{fig:maxclique} we present the mean values of this analysis.
The plot shows that the greater the subgraph, the greater the value of this metric.
Nevertheless the maximum clique number has a small declining trend for the same number of sampled graph over time. 
This trend is more obvious for the per-month subplot that since it contained fewer nodes allowed us to test for more sampling sizes.
This trend shows that the graph while it grows becomes more sparse in strongly connected components.

\subsection{Closeness}
\label{sec:closeness}
\begin{figure}[t]
  \centering
  \includegraphics[width=0.48\textwidth]{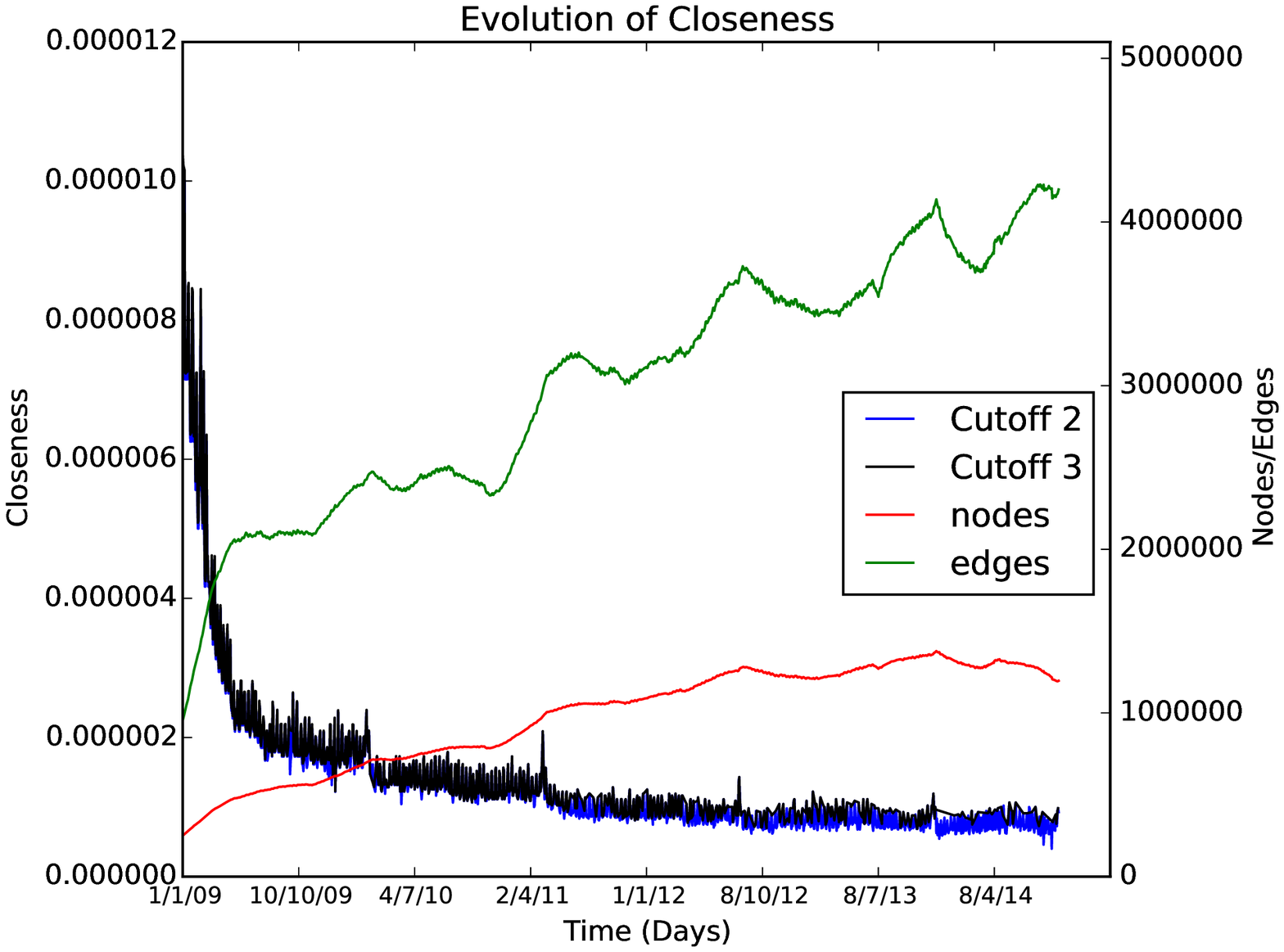}\llap{\makebox[11.6cm]{\raisebox{3.95cm}{\includegraphics[height=2cm]{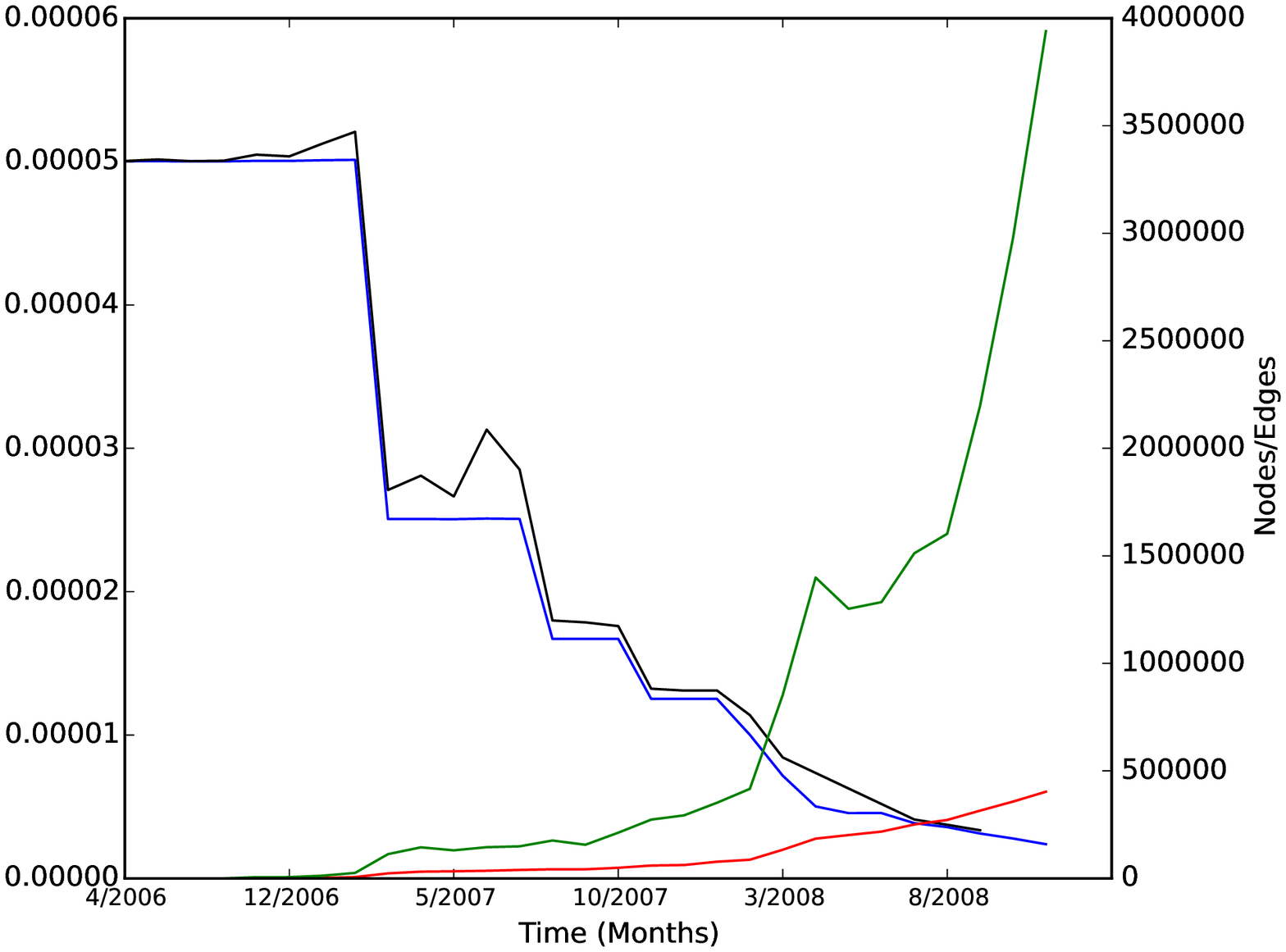}}}}
  \caption{Closeness per day from start of 2009 to the end of 2014 and per month from April 2006 to December 2009 (embedded graph).}
  \label{fig:closeness}
\end{figure}

Closeness of a node is the inverse of the sum of the lengths of all geodesics from or to the given node. 
A geodesic is a shortest path between two nodes. 
Closeness gives an indication of how easy is to reach other nodes from this node.
Equivalently it measures how easy is to access this node from other nodes. 
Figure~\ref{fig:closeness} presents a steady decrease of closeness over time.
This means that the graph becomes more compact while it grows over time making easier the access of a node from any other node. 
In this plot we measure the average closeness for all metrics for two cutoff values: 2 and 3. 
A cutoff means that only paths of this length are considered when estimating this measure.
The figure shows that these two cutoff values return approximately equivalent results. 

\subsection{Cocitation}
\label{sec:cocitation}
\begin{figure}[t]
  \centering
  \includegraphics[width=0.48\textwidth]{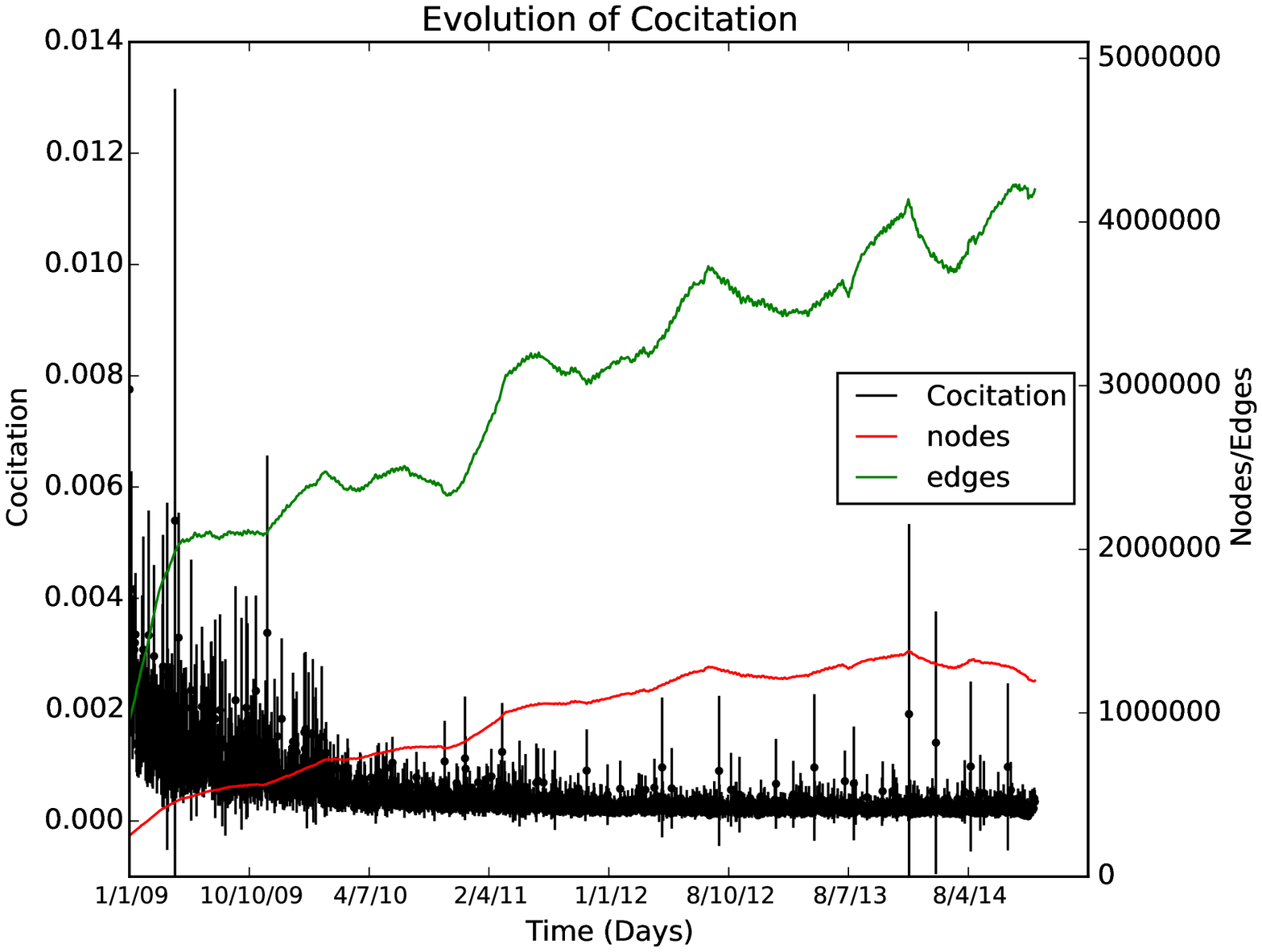}\llap{\makebox[12.0cm]{\raisebox{3.95cm}{\includegraphics[height=2cm]{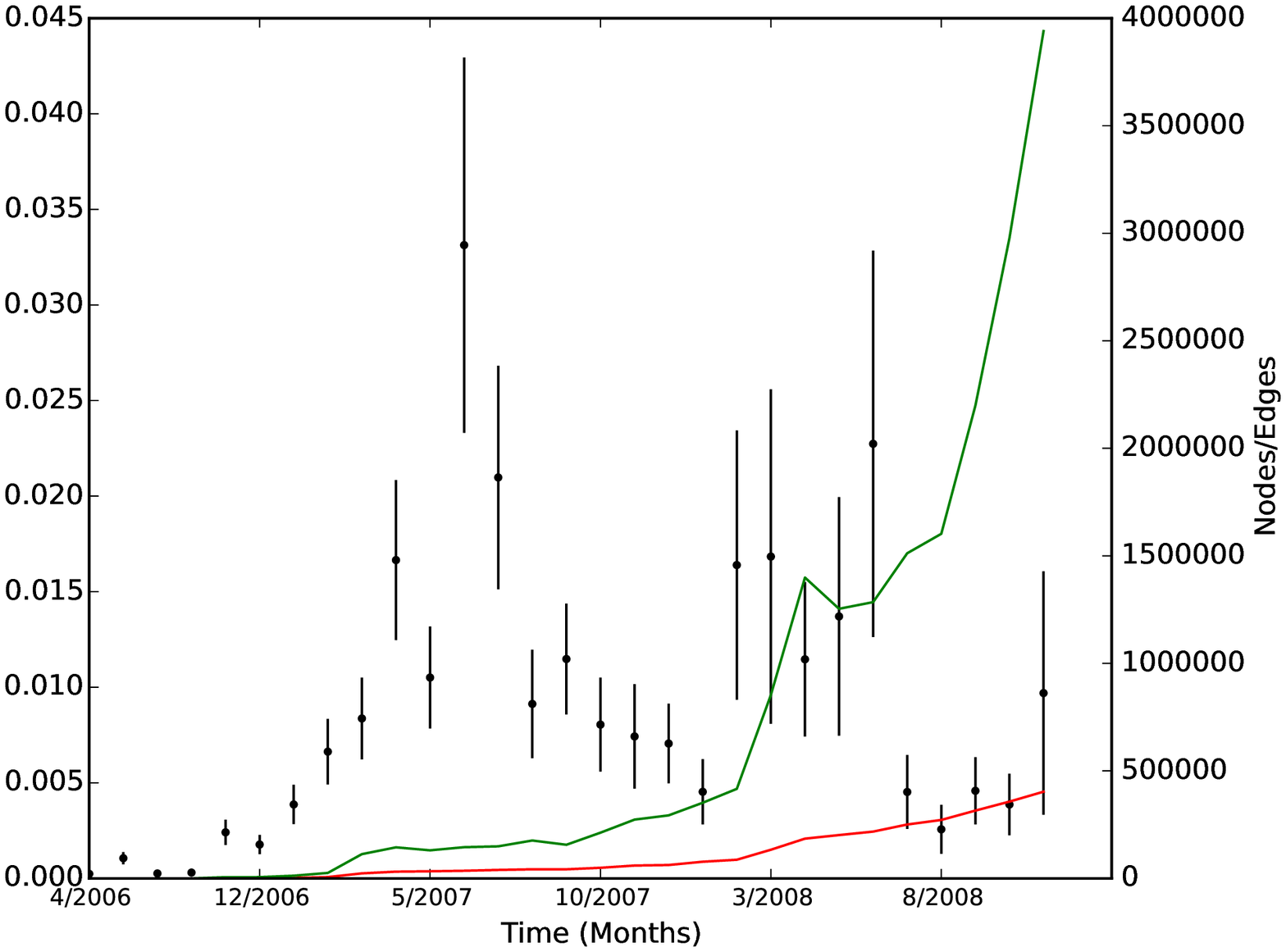}}}}
  \caption{Cocitation per day from start of 2009 to the end of 2014 and per month from April 2006 to December 2009 (embedded graph). The black vertical lines show the 95\% confidence intervals of the sampling procedure.}
  \label{fig:cocitation}
\end{figure}

Two nodes are cocited if there exist at least one node that `cites' both of them. 
By `cite' here we mean connect with a single edge. 
A citation score between two nodes is the number of other nodes that are directly connected to both of them.
This metric measures for each node the cocitation score with every other node. 
Thus it returns a two dimensional list (or else, a list of lists).
In order to report a single value for the complete graph we measure the mean of the mean scores for each node.
Moreover due to the time complexity required in this metric, we applied the Random Node sub-sampling presented in section ~\ref{sec:measurementmethods}.
On Figure~\ref{fig:cocitation} we present the results which contain the mean values from the sampling along with the 95\% confidence intervals (vertical lines). 
We notice that at the beginning there is a high uncertainty and deviation of values but after 2010 the values are stabilized closed to zero.
It is interesting that although the graph increases over time the cocitation score is not affected and remains close to zero. 

\subsection{Average Degree}
\label{sec:averagedegree}
\begin{figure}[t]
  \centering
  \includegraphics[width=0.48\textwidth]{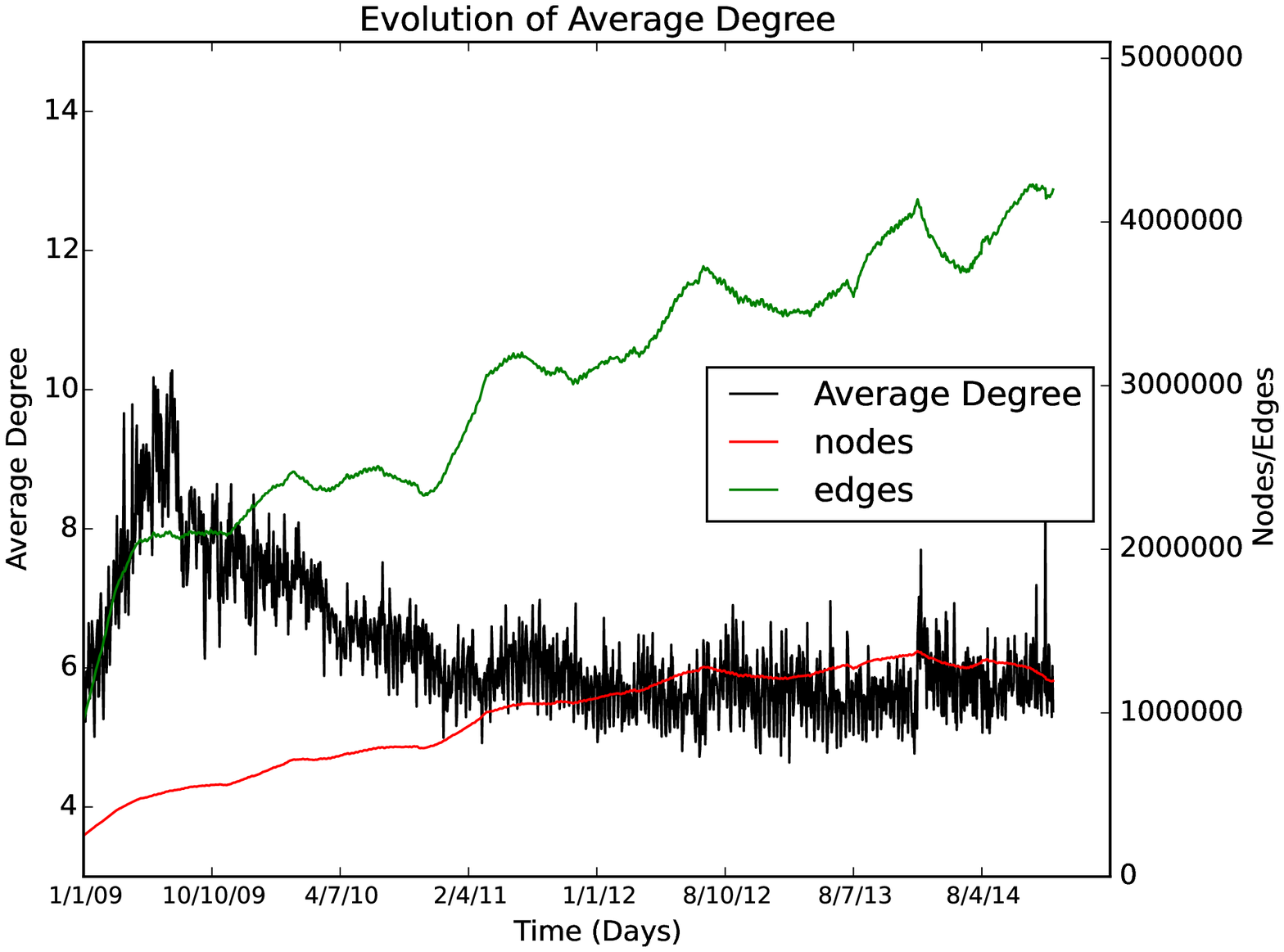}\llap{\makebox[12.5cm]{\raisebox{3.95cm}{\includegraphics[height=2cm]{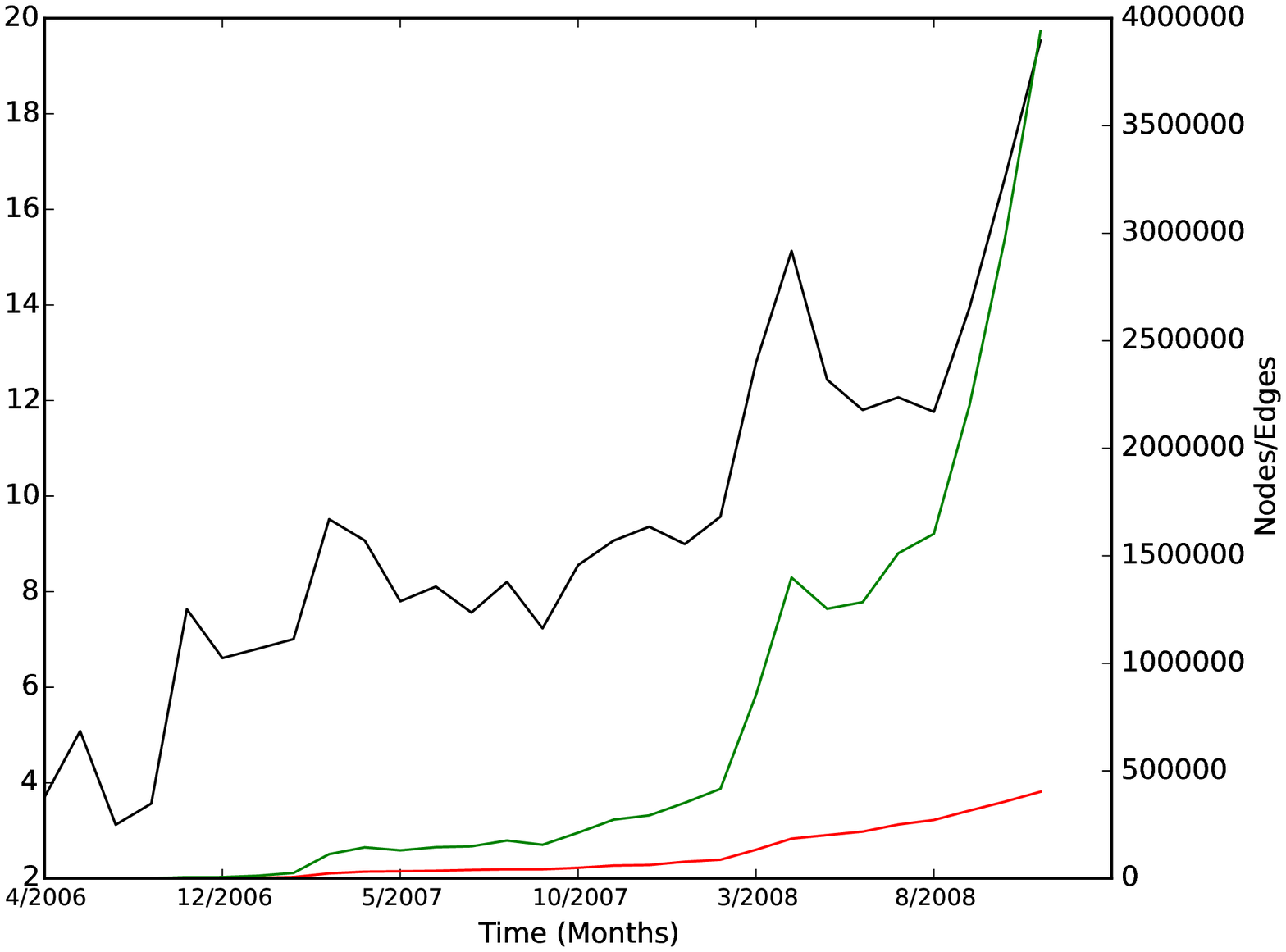}}}}
  \caption{Average degrees per day from start of 2009 to the end of 2014 and per month from April 2006 to December 2009 (embedded graph).}
  \label{fig:degree}
\end{figure}

The average degree is one of the most studied properties of Twitter \cite{Kwak2010}) and OSNs in general \cite{leskovec2007graph}. 
It is a well established fact that the degree distribution is indicative of a scale-free structure. 
Here, on Figure \ref{fig:degree} we plot the evolution of the average degree in our dataset.
It is evident that Twitter has been gone through many growth periods.
From the beginning (April 2006) until the middle of 2009 Twitter experiences a very rapid growth and it seems that nodes, edges and consequently the average degree follow the same growth rates.
From the middle of 2009 until the start of 2011 although the nodes and edges continue to grow, the average degree seems to follow a `correction' course and drops to 6.  
After that the average degree stabilizes close to 6 and remains at this point until the end of 2014.
It is essential that during these periods the nodes and edges show small variability in their growth rates. 

\subsection{Coreness}
\label{sec:coreness}
\begin{figure}[t]
  \centering
  \includegraphics[width=0.48\textwidth]{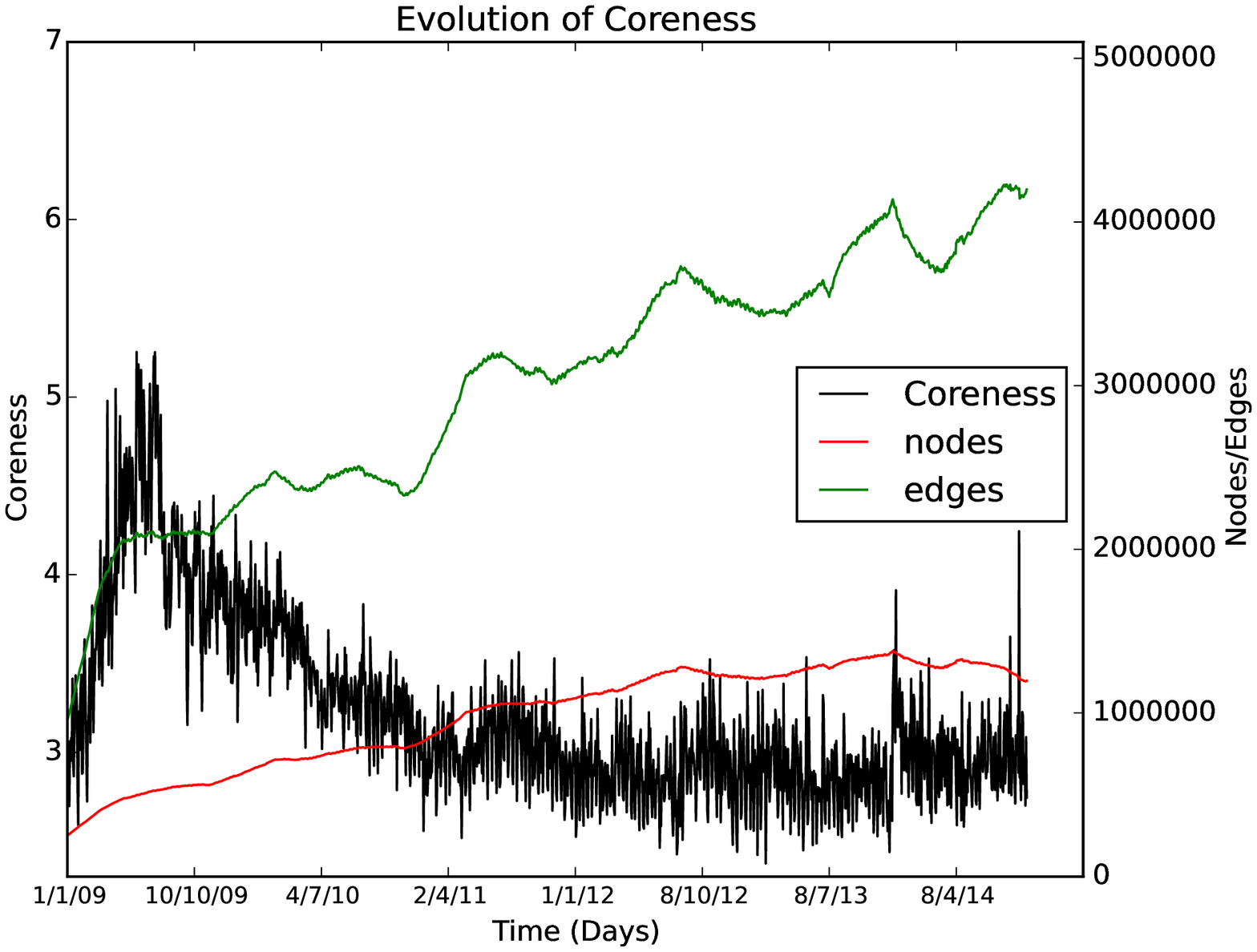}\llap{\makebox[11.5cm]{\raisebox{3.95cm}{\includegraphics[height=2cm]{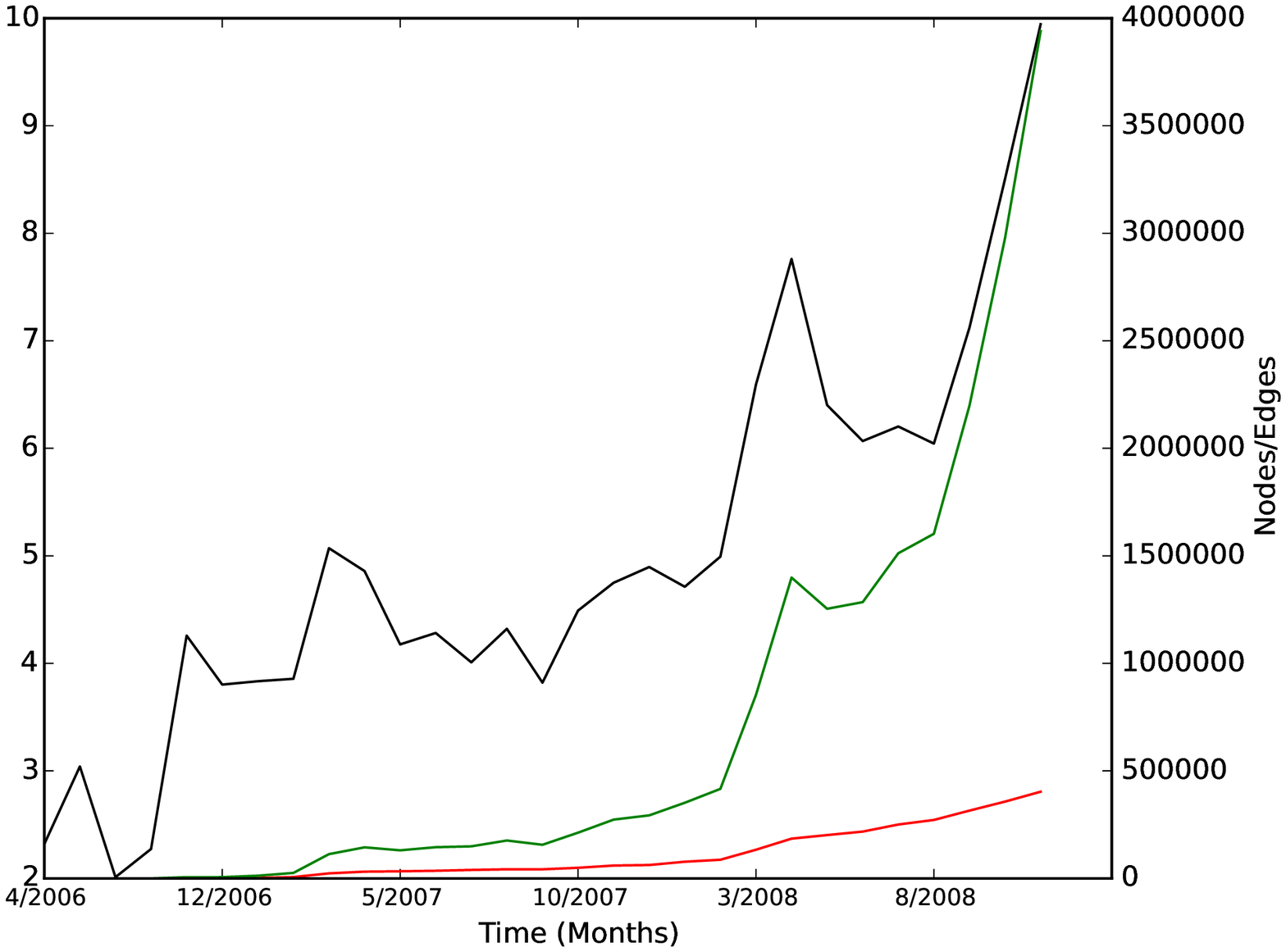}}}}
  \caption{Coreness per day from start of 2009 to the end of 2014 and per month from April 2006 to December 2009 (embedded graph).}
  \label{fig:coreness}
\end{figure}

The Coreness (or shell index) of a node is a measure of the compactness of it's surrounding neighborhood. 
If the coreness of a node is k then there exist a subgraph containing this node where each node has a degree of at least k (but it does not exist a subgraph where each node has a degree of k+1)~\cite{batagelj2003m}.
The figure~\ref{fig:coreness} presents the evolution of average coreness over all the nodes of the graph.
There is high similarity between the coreness evolution with the evolution of the Average Degree that presented in subsection~\ref{sec:averagedegree}.
This illustrates the effect of changes of the nodes degree over time, on the structure of small communities in the graph.

\subsection{Diameter}
\label{sec:diameter}
\begin{figure}[t]
  \centering
  \includegraphics[width=0.48\textwidth]{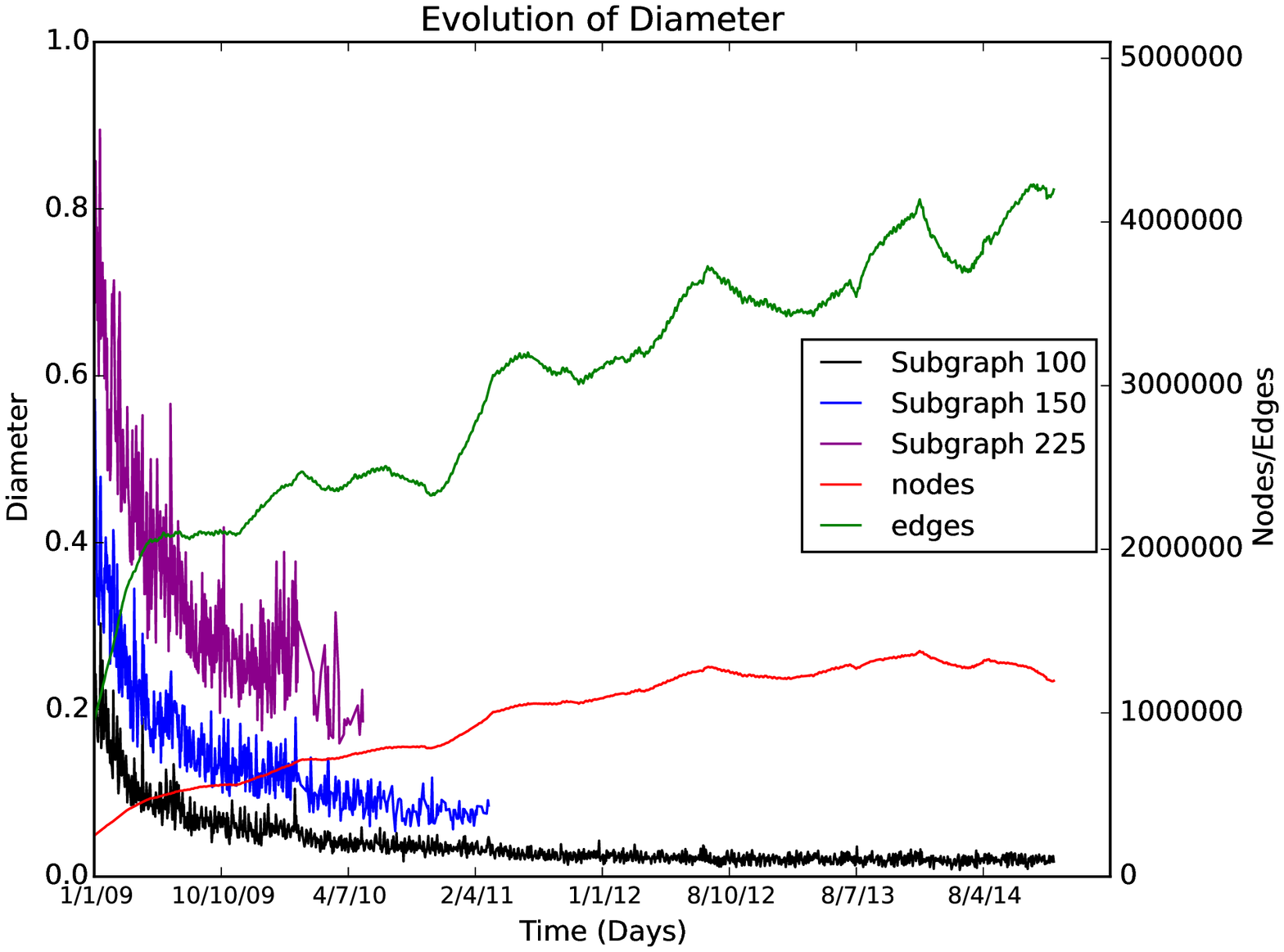}\llap{\makebox[12.5cm]{\raisebox{3.95cm}{\includegraphics[height=2cm]{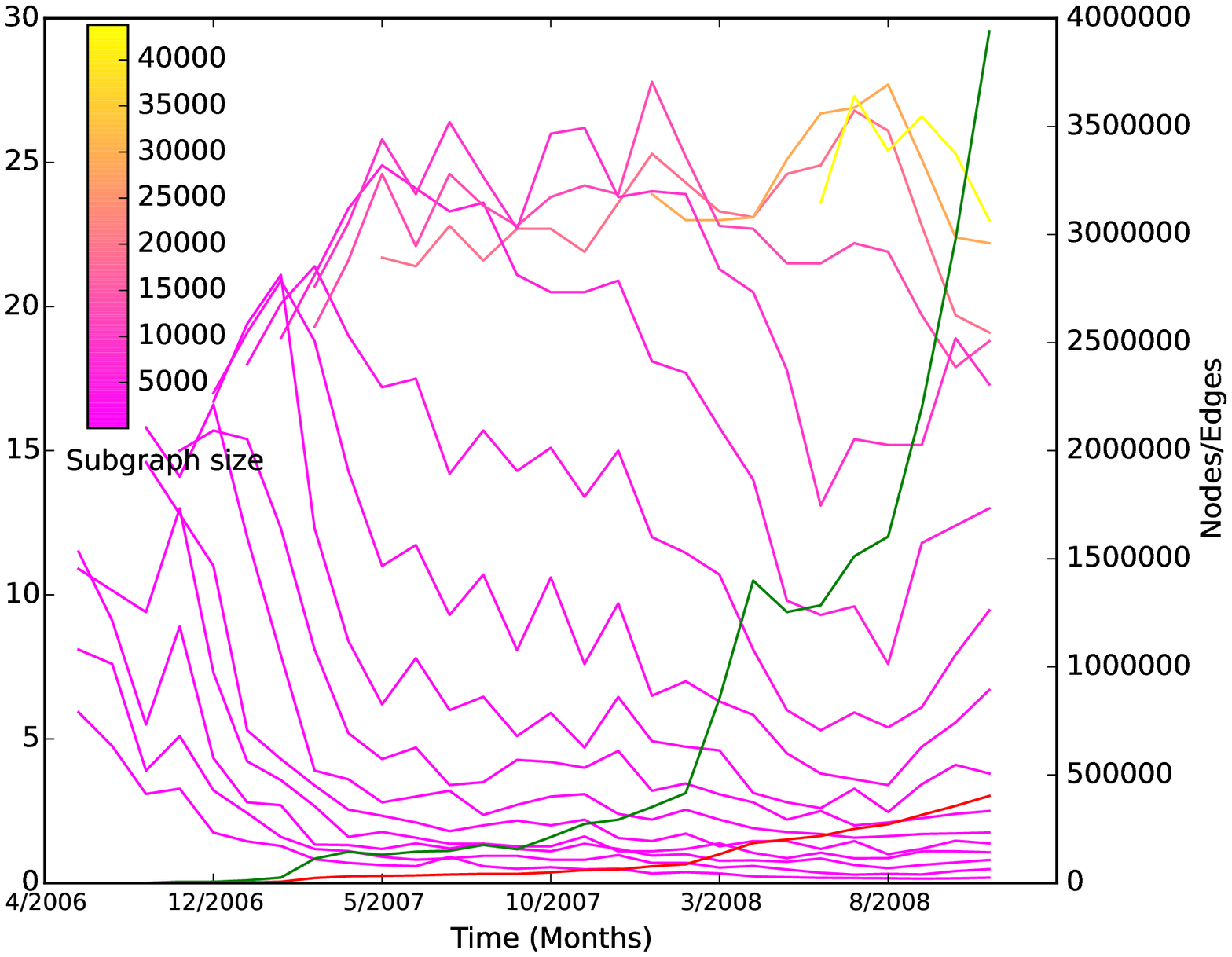}}}}
  \caption{Diameter per day from start of 2009 to the end of 2014. Measurement for random subgraphs of size 150 and 225 was partially estimated. The embedded graph shows the months from April 2006 to December 2009.}
  \label{fig:diameter}
\end{figure}

The diameter of a graph is the longest shortest path between any two nodes of the graph.
This measure requires computational time proportional to the number of nodes multiplied by the number of edges of the graph.
For this reason we applied the subgraph sampling technique which was able to infer values for random subgraphs of 100 nodes.
Values for random subgraphs of 150 and 225 nodes were inferred for the initial days of our dataset.
On Figure~\ref{fig:diameter} we can see a downward trend of the diameter on these small subgraphs. 
These figures indicate that isolated nodes are becoming fewer and the network becomes more dense. 

\subsection{Density}
\label{sec:density}
\begin{figure}[t]
  \centering
  \includegraphics[width=0.48\textwidth]{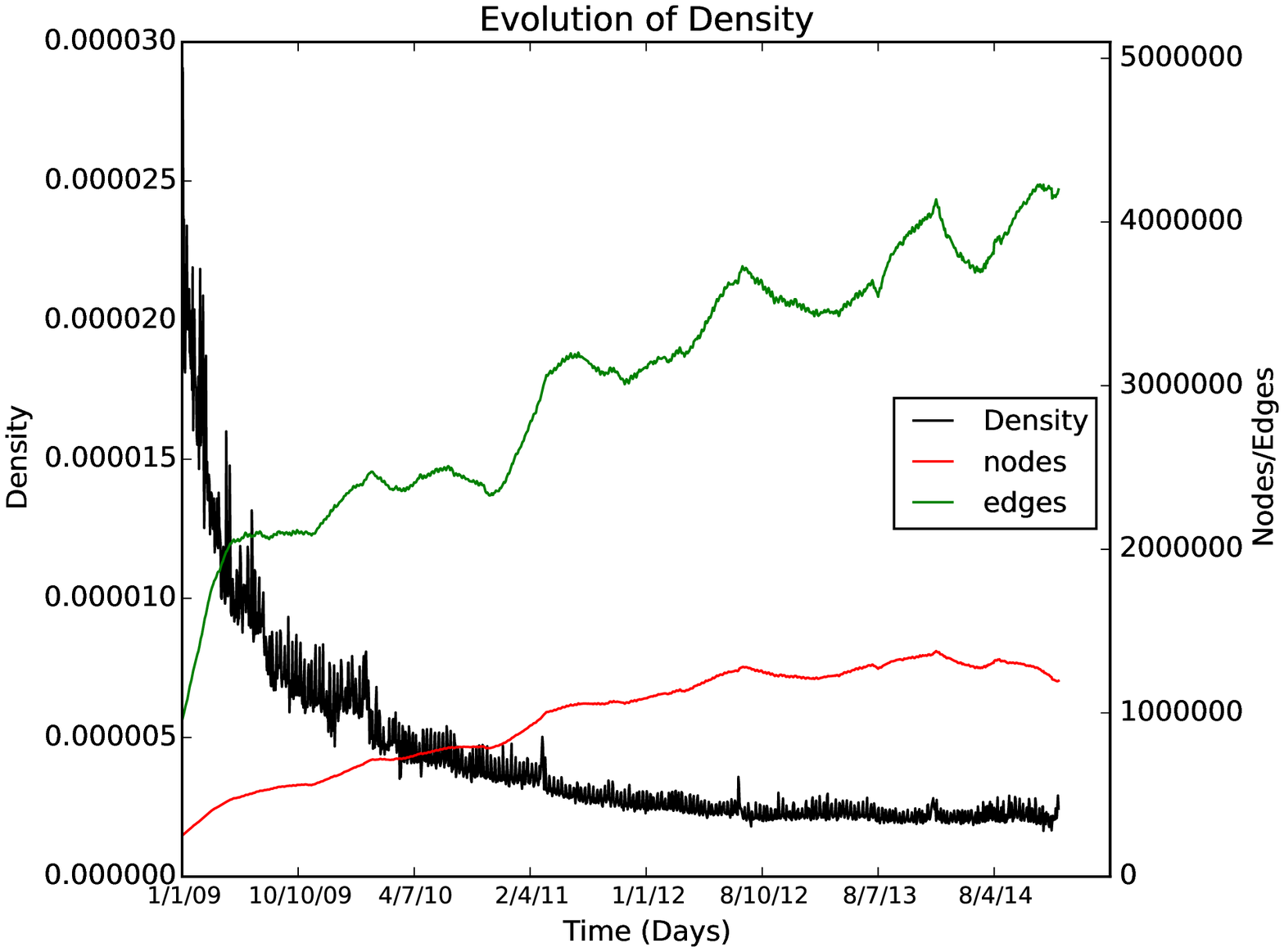}\llap{\makebox[11.5cm]{\raisebox{3.95cm}{\includegraphics[height=2cm]{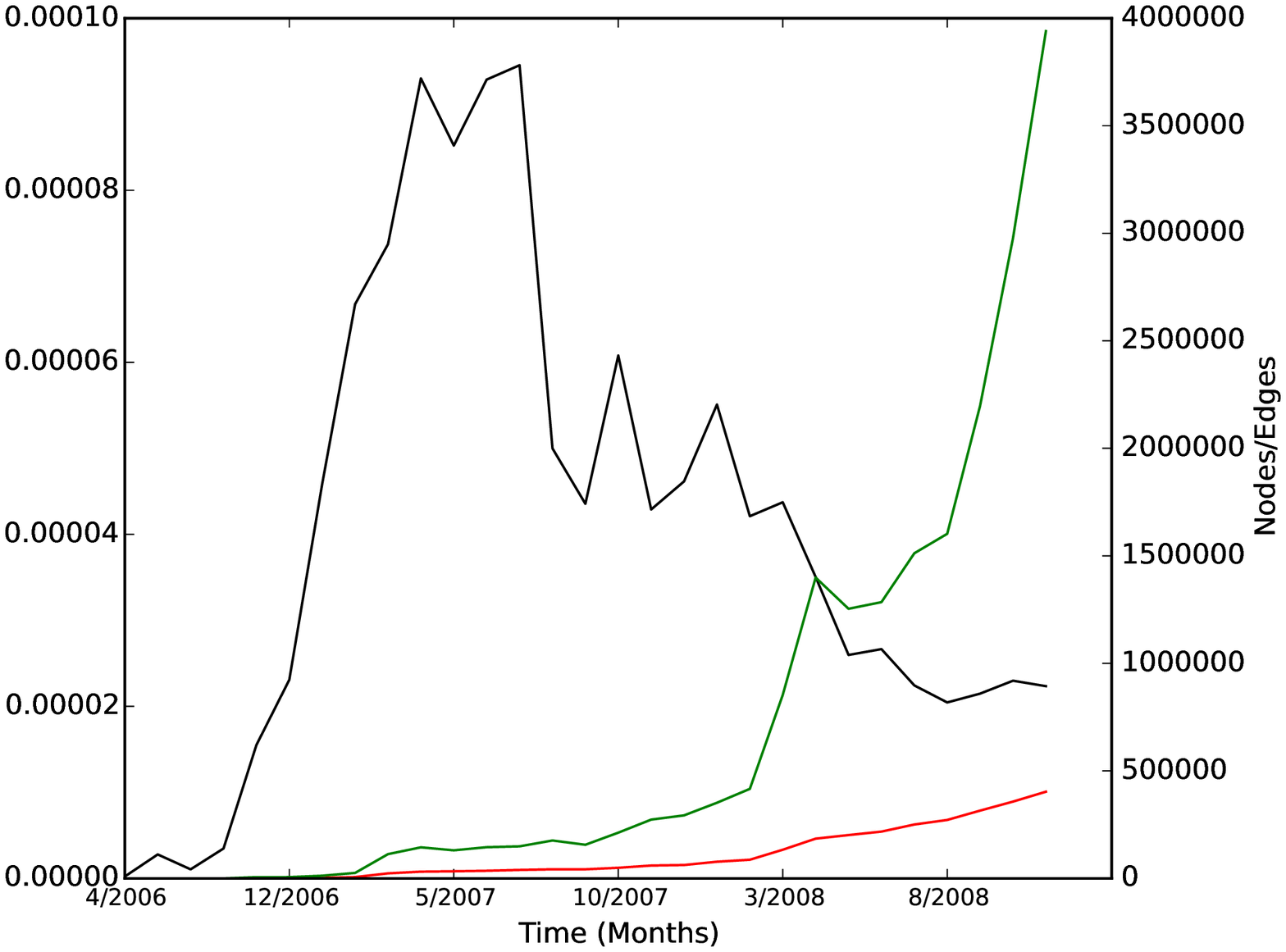}}}}
  \caption{Density per day from start of 2009 to the end of 2014 and per month from April 2006 to December 2009 (embedded graph).}
  \label{fig:density}
\end{figure}

Density is a measure that shows how close is the number of edges of a graph to the maximum number of edges.
For a directed graph, like in our case, this metric is defined as:
$$D = \frac{|E|}{|V|\,(|V|-1)}$$
From Figure~\ref{fig:density} is evident that the density of the graph drops throughout time. 
Although the edges grow in a much higher degree than nodes, the addition of new nodes expands the space of possible edges in a quadratic to the number of nodes rate.

\subsection{Eccentricity}
\label{sec:eccentricity}
\begin{figure}[t]
  \centering
  \includegraphics[width=0.48\textwidth]{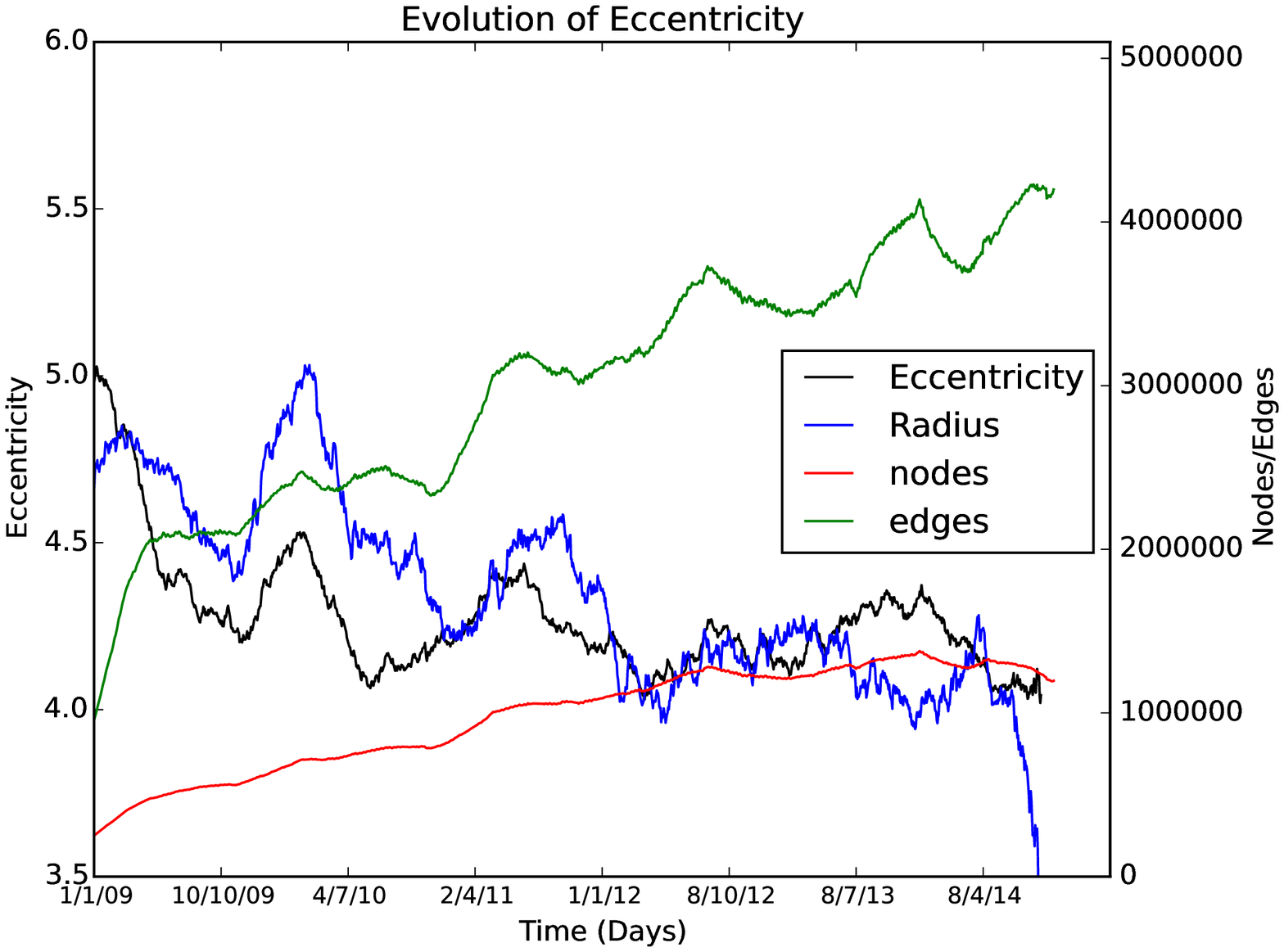}\llap{\makebox[12.0cm]{\raisebox{3.95cm}{\includegraphics[height=2cm]{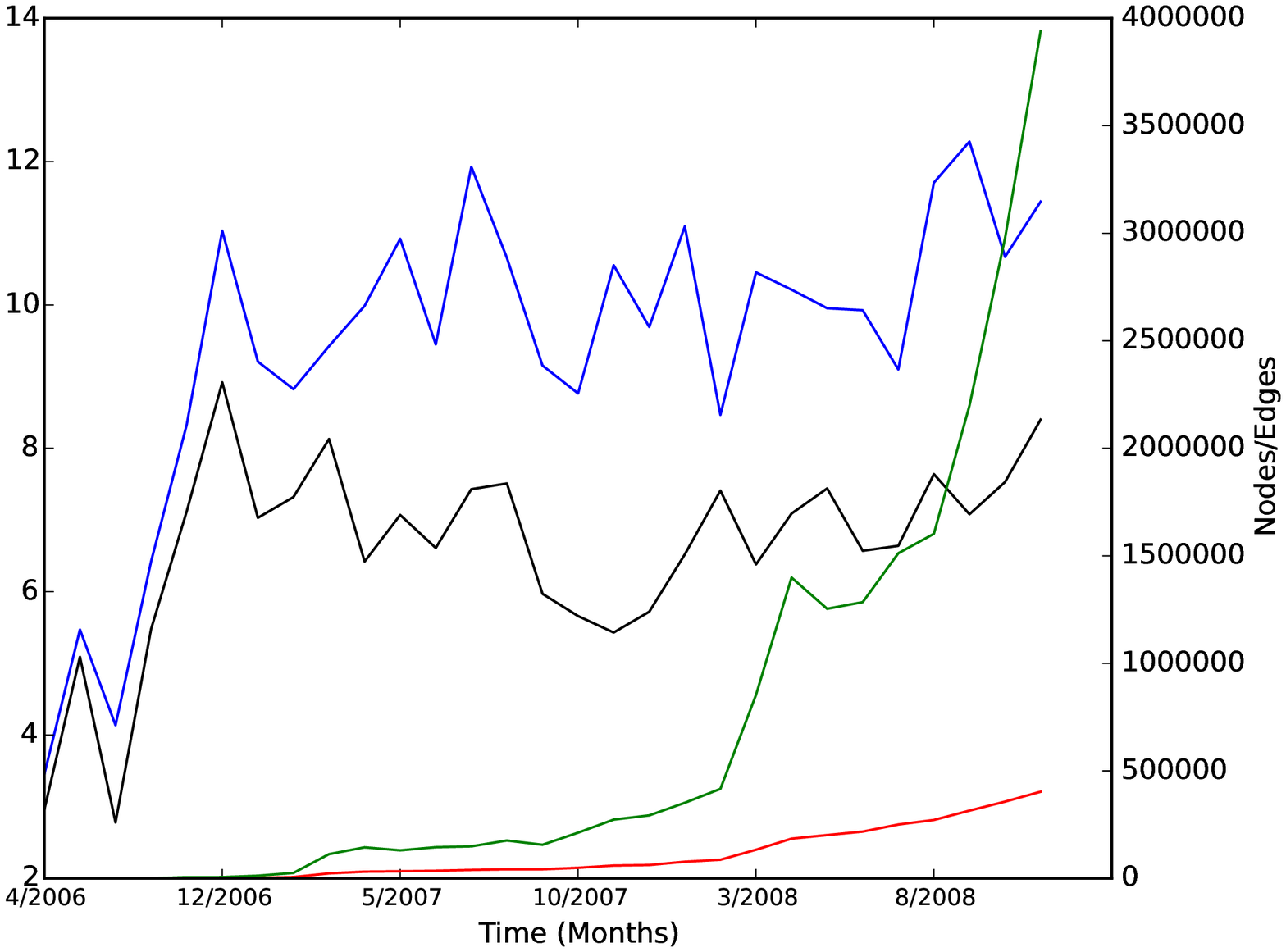}}}}
  \caption{Average Eccentricity and Radius per day from start of 2009 to the end of 2014 and per month from April 2006 to December 2009 (embedded graph). The values of radius is not shown on the graph.}
  \label{fig:eccentricity}
\end{figure}

Eccentricity measures how distant a node is compared to the rest of the nodes in the graph.
It is equal to the maximum shortest distance between this node and every other node in the graph.
Here we applied the Random Nodes graph sampling technique. 
Another interesting measurement is the `radius' which is the minimum eccentricity of the graph.
On figure~\ref{fig:eccentricity} we present the average eccentricity of the Random Nodes along with the radius for all time points.
The graph shows a small and fluctuating downward trend for both values as of the end of 2008.
This drop is an evidence that while the graph grows fewer nodes remain isolated. 
It is also a sign of a decrease of the sparseness of the graph.

\subsection{Eigenvector Centrality}
\label{sec:eigenvectorcentrality}
\begin{figure}[t]
  \centering
  \includegraphics[width=0.48\textwidth]{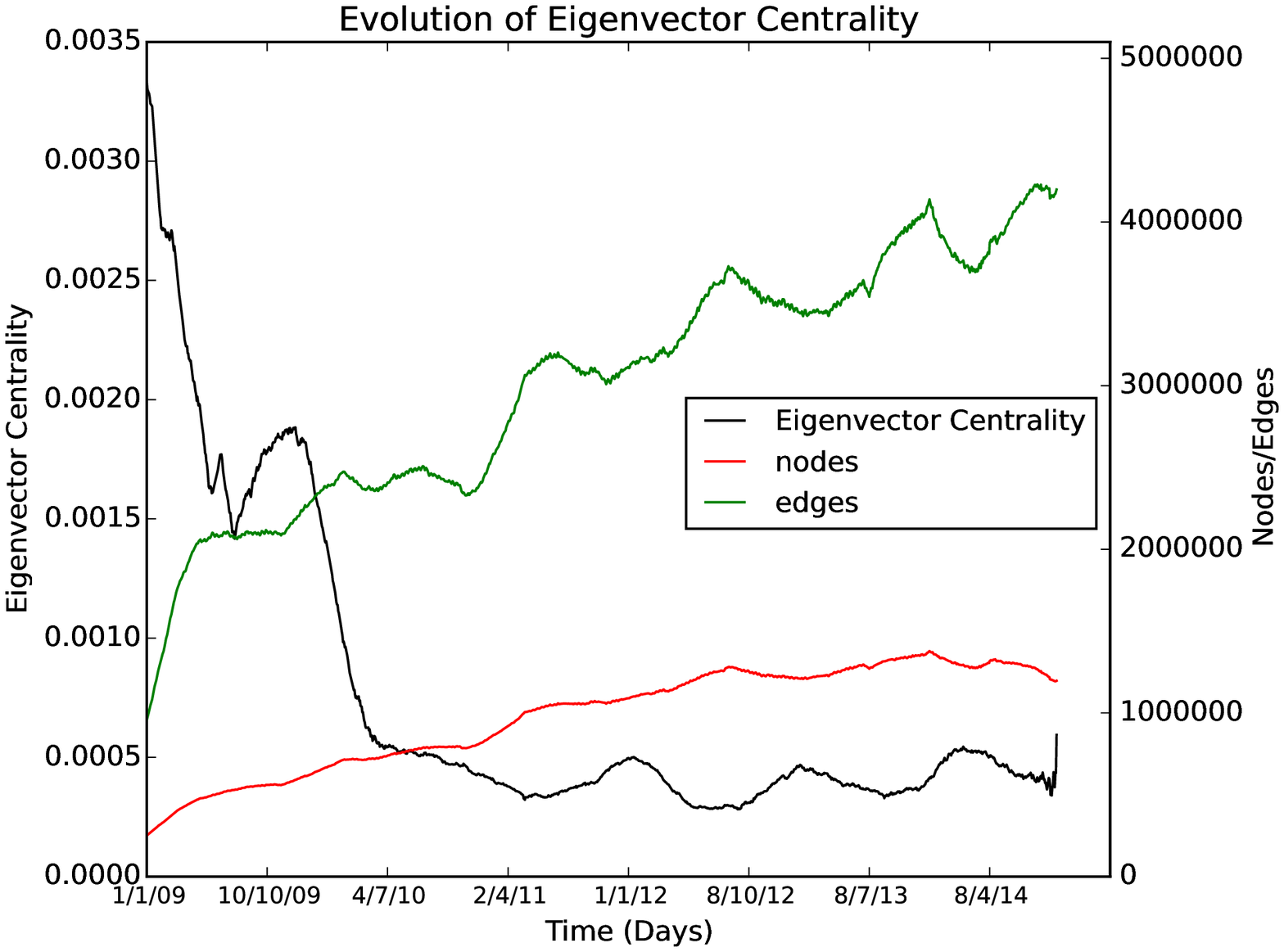}\llap{\makebox[11.5cm]{\raisebox{3.95cm}{\includegraphics[height=2cm]{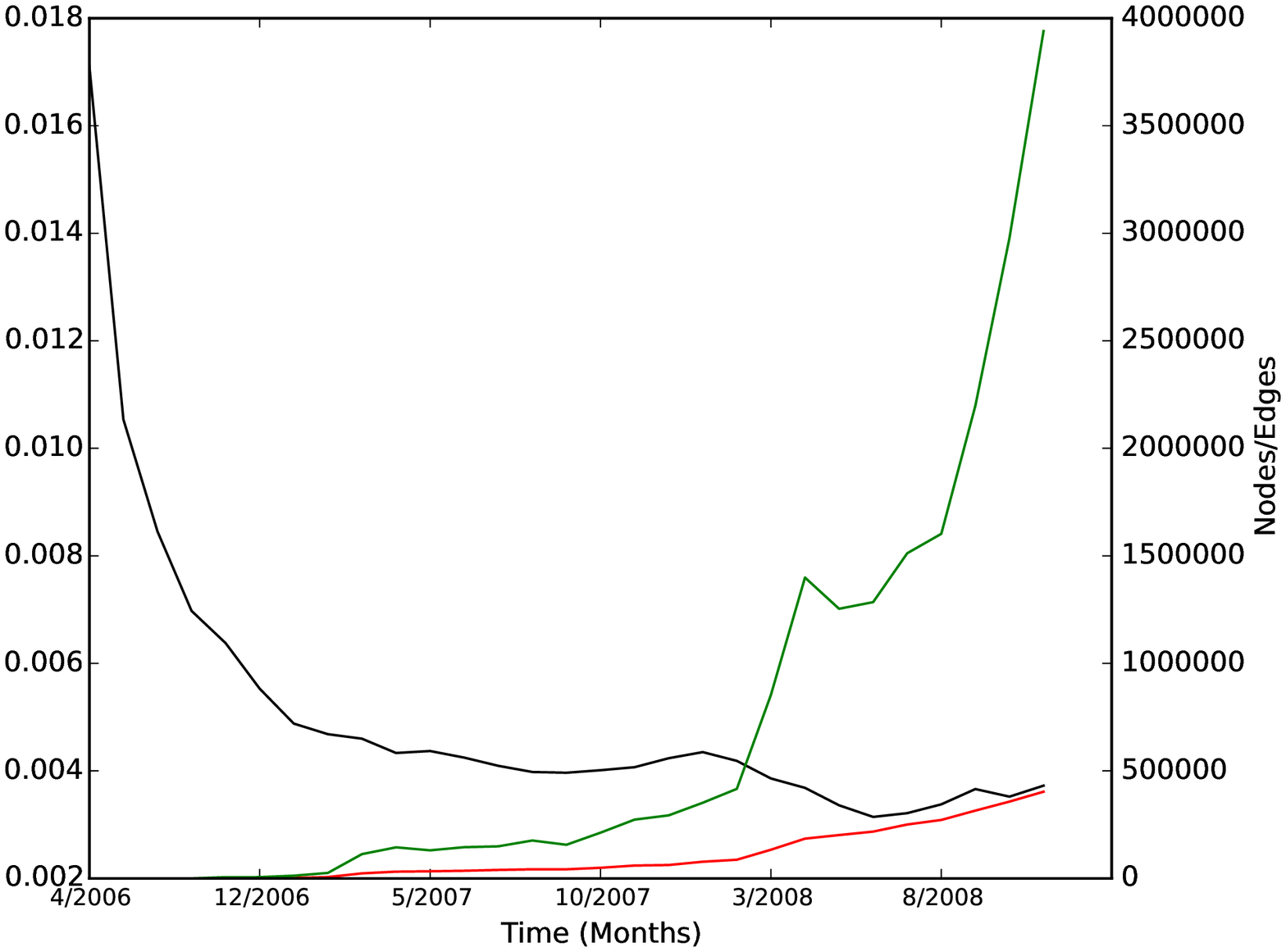}}}}
  \caption{Average Eigenvector Centrality per day from start of 2009 to the end of 2014 and per month from April 2006 to December 2009 (embedded graph).}
  \label{fig:eigenvectorcentrality}
\end{figure}

This metric measures the influence of a node in the graph. 
It is based on the idea that the influence of a node is increased if it is connected to a node that is itself influential (and decreased if it is not influential). 
So the influence of a node can be defined as the average of the influences of the nodes that is connected.
This can be formed as an eigenvector equation and solved with linear algebra methods~\cite{newman2008mathematics}.
Figure~\ref{fig:eigenvectorcentrality} shows the average eigenvector centrality throughout time. 
The first observation is that the value of the measure starts dropping from Twitter's creation time until 2010 where it stabilizes.
We can also notice that the values of this metric can fluctuate depending on the number of edges on the graph. 
Random fluctuations on the number of edges on our dataset seem to be inverse associated to the values of this metric.

\subsection{Average Shortest Path}
\label{sec:eigenvectorcentrality}
\begin{figure}[t]
  \centering
  \includegraphics[width=0.48\textwidth]{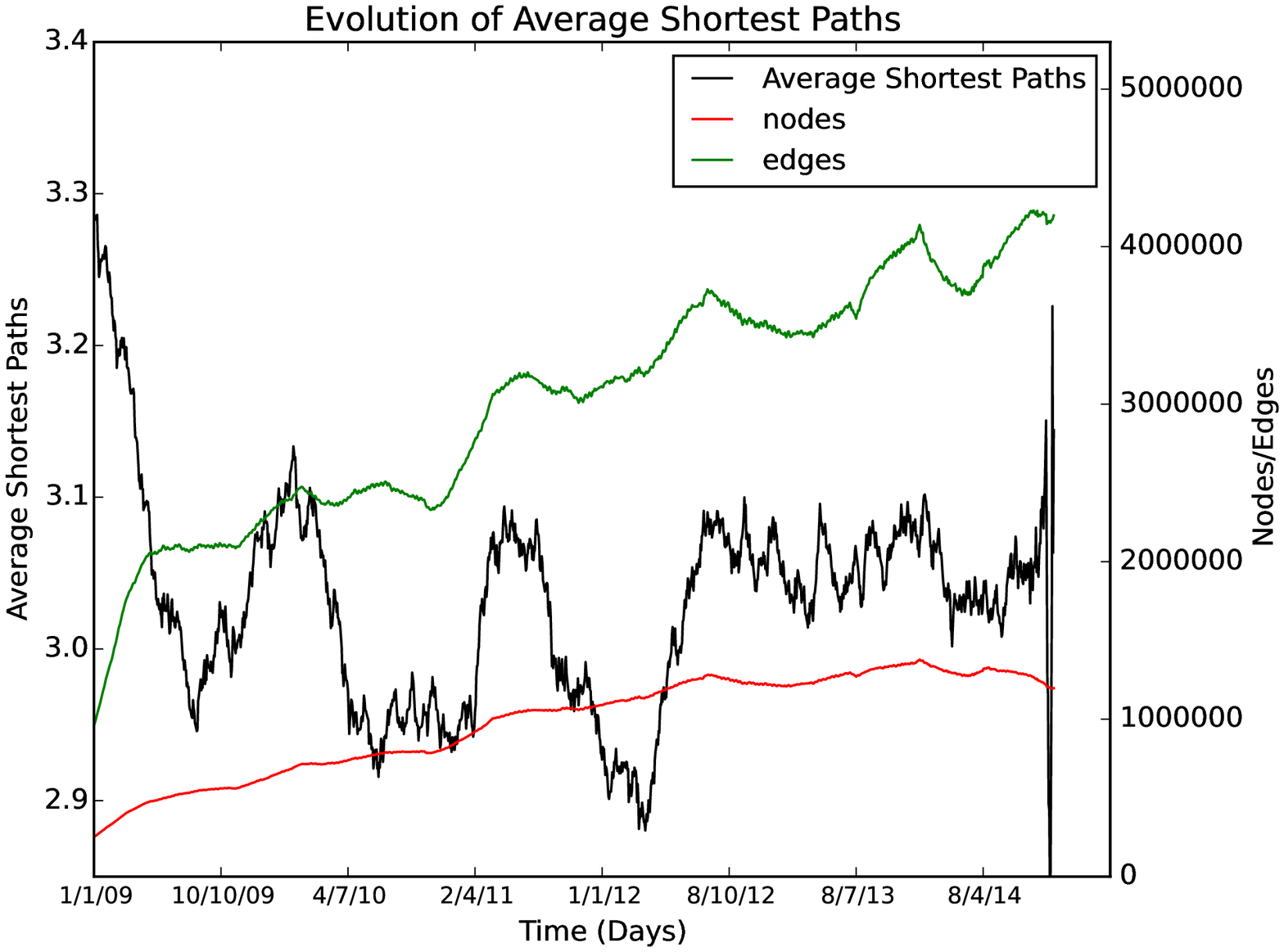}\llap{\makebox[12.0cm]{\raisebox{3.95cm}{\includegraphics[height=2cm]{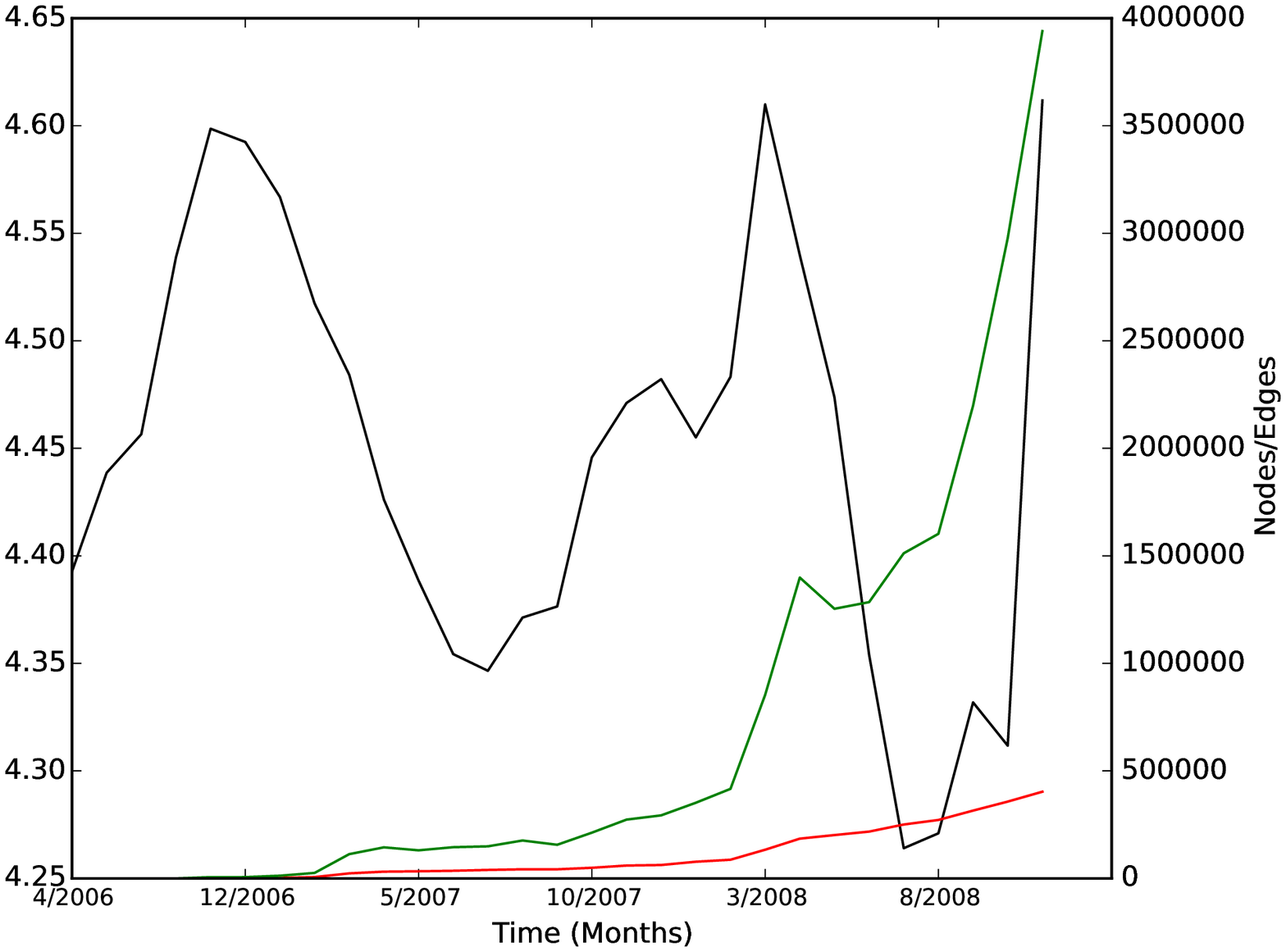}}}}
  \caption{Average Shortest Path per day from start of 2009 to the end of 2014 and per month from April 2006 to December 2009 (embedded graph).}
  \label{fig:shortestpaths}
\end{figure}

This is one of the most well known graph metrics that shows the general sparseness of the OSN.
In our metrics we applied the Random Nodes technique in order to get an approximation of the shortest average path for the various time points in our dataset.
The same technique was used by \cite{Kwak2010}. 
On figure~\ref{fig:shortestpaths} we notice that at the early stages of Twitter, the average shortest path is higher and varies on values close to 4.4.
After 2009 the value drops and fluctuates between 2.9 and 3.1.
The average shortest path seems to be independent from the growth of nodes and edges in the graph after 2009.
The value of this metric has been associated with the six-degrees of separation. 
Recent studies has shown that in OSNs the average shortest path is lower. 
For example \cite{Backstrom2012} demonstrated that 4 is closer to the real value of the average number of the intermediates on a random shortest path between two nodes on the Facebook OSN. 
The even lower values that we report can be attributed on the per day splitting of our dataset. 
We expect graphs that contain longer periods to have higher average shortest paths but not higher than 4.

\subsection{Kleinberg's hub score}
\label{sec:hubscore}
\begin{figure}[t]
  \centering
  \includegraphics[width=0.48\textwidth]{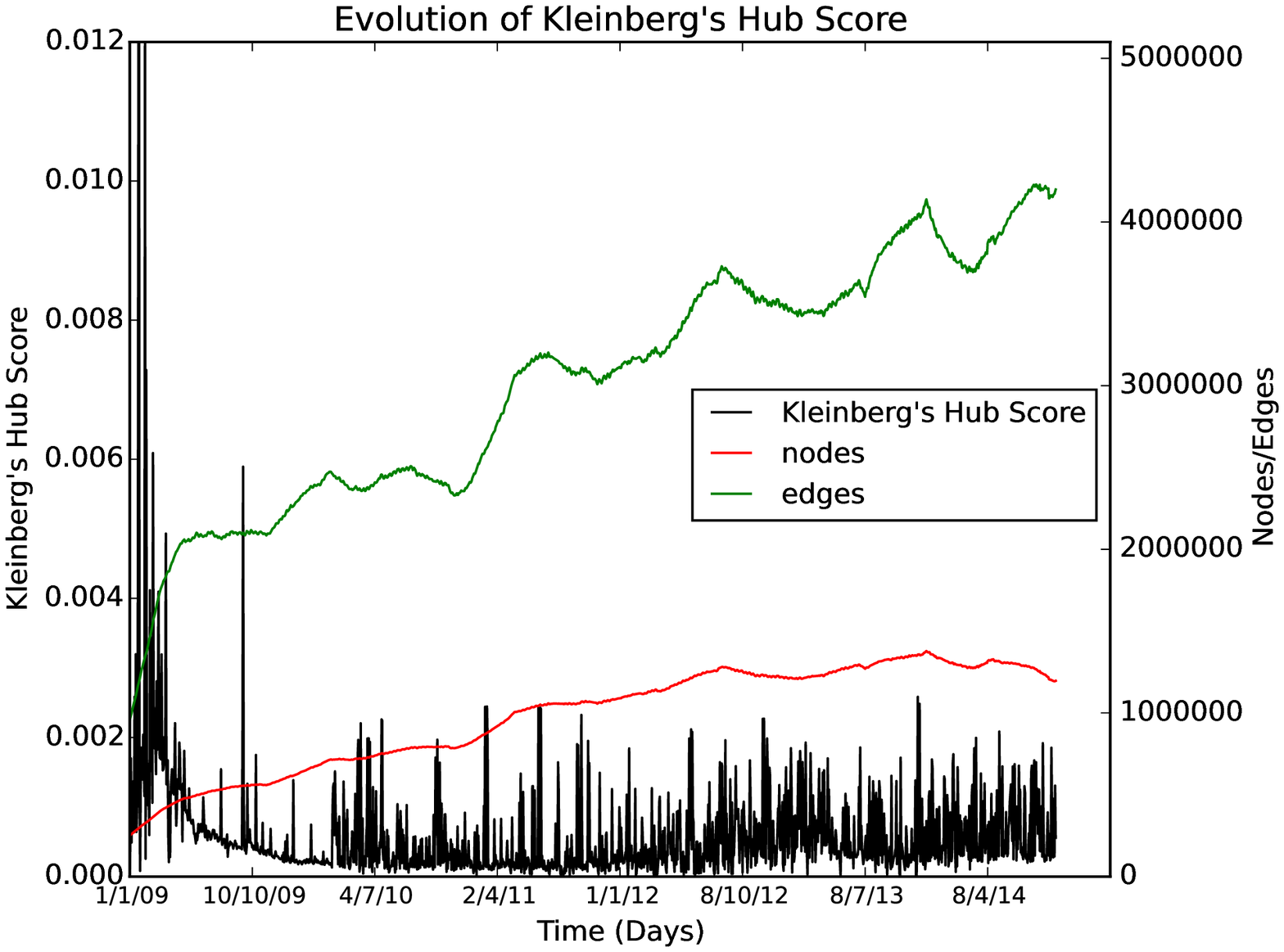}\llap{\makebox[12.0cm]{\raisebox{3.95cm}{\includegraphics[height=2cm]{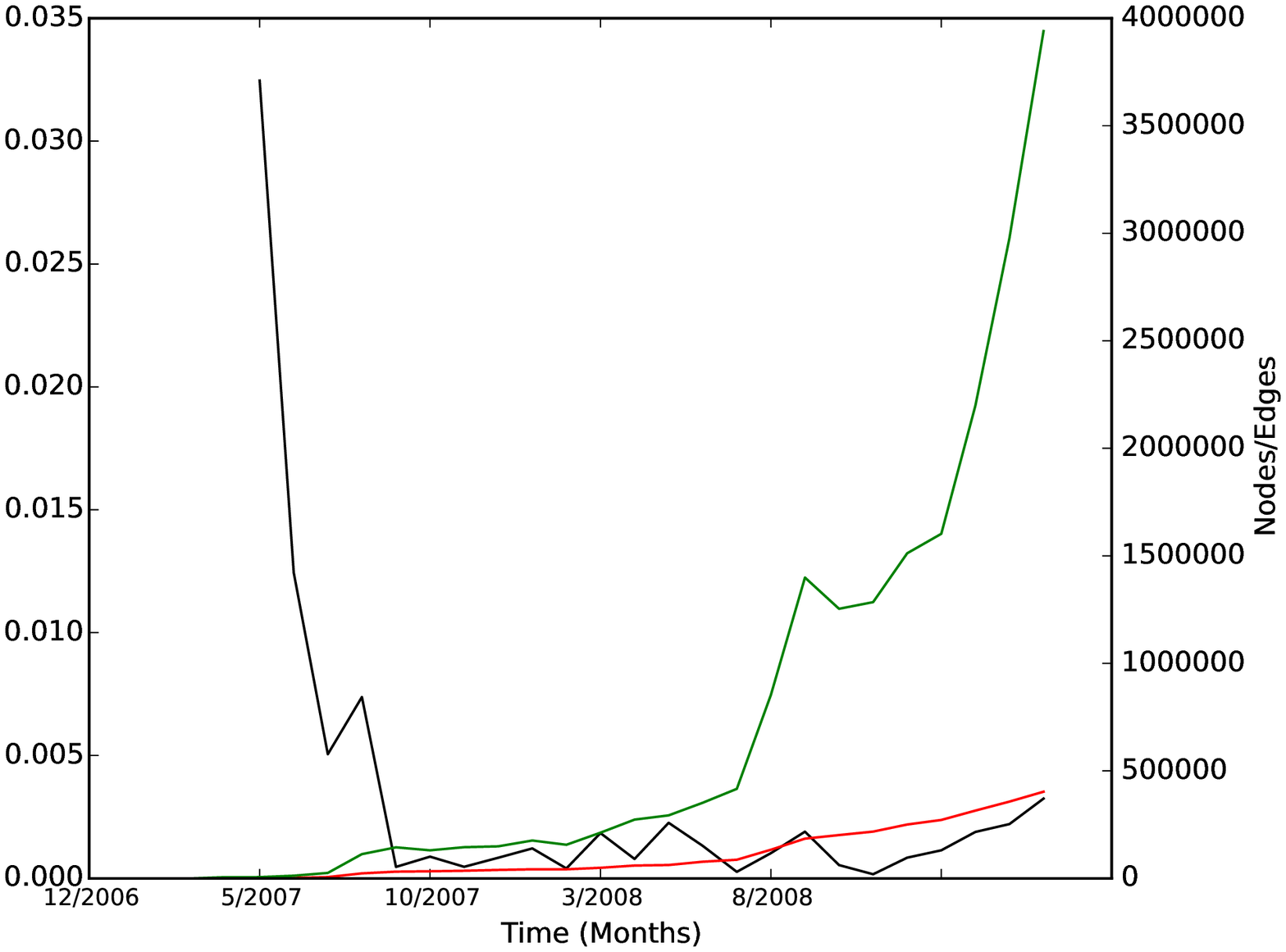}}}}
  \caption{Average Kleinberg's Hub Score per day from start of 2009 to the end of 2014 and per month from April 2006 to December 2009 (embedded graph).}
  \label{fig:hubscore}
\end{figure}

Kleinberg's hub score \cite{kleinberg1999web} assigns two values on each node: the Hub score and the Authority score.
Nodes with high Hub score, have high out-degree and act as information flow gateways.
On the other side, nodes with high Authority score have a high in-degree and are common ending points of this information.
The Out- and In- degree of a node in a directed graph is the number of outgoing and ingoing to this node respectively.
This metric was first used to measure the influence of web pages mainly at the early stages of WWW. 
On figure~\ref{fig:hubscore} we plot the average Kleinberg's Hub Score of every node for every time point.
Twitter's OSN exhibited a high hub score at the beginning that topped at the middle of 2009. 
After that it stabilized with small fluctuations on values little higher than zero.

\subsection{Neighbors Average Degree}
\label{sec:knnaveragedegree}
\begin{figure}[t]
  \centering
  \includegraphics[width=0.48\textwidth]{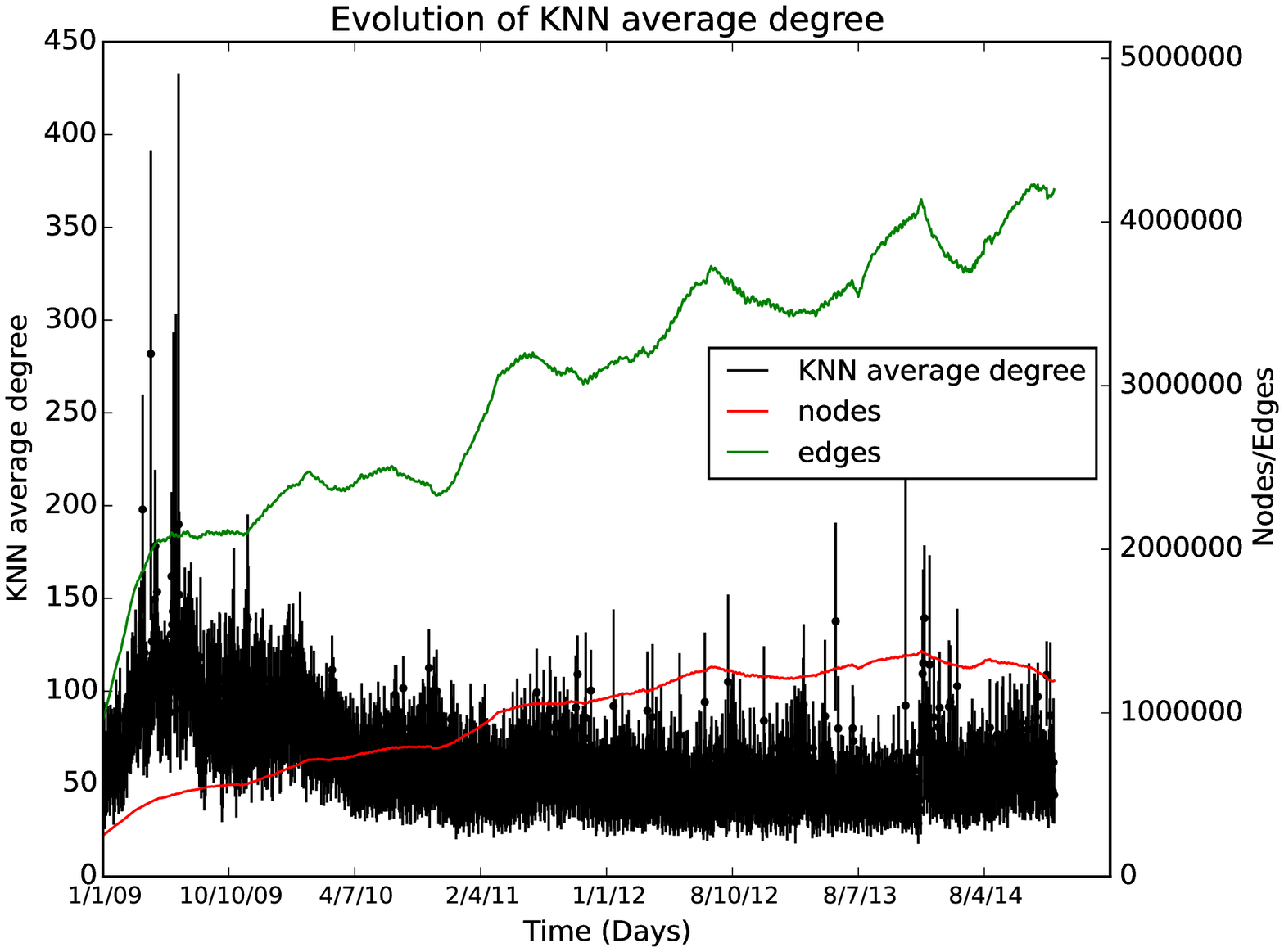}\llap{\makebox[12.0cm]{\raisebox{3.95cm}{\includegraphics[height=2cm]{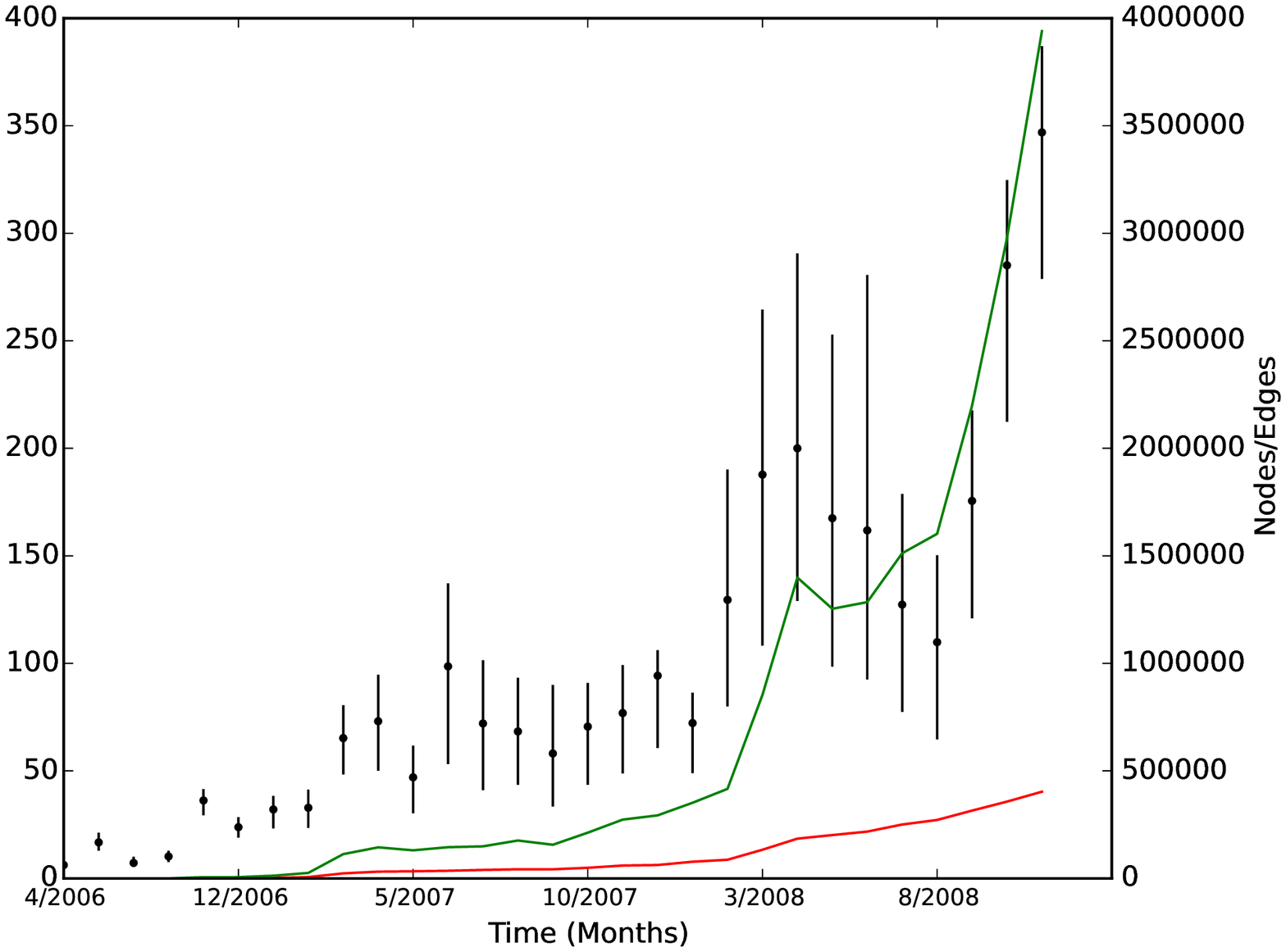}}}}
  \caption{Boxplots of Neighbors Average Degree per day from start of 2009 to the end of 2014 and per month from April 2006 to December 2009 (embedded graph).}
  \label{fig:knn}
\end{figure}

This metric calculates for each node the average degree of the nodes that is connected to (or else the Nearest Neighbors)~\cite{Barrat2004}.
This metric belongs to the `architecture' family of measures since it measures the overall connectivity.
It is interesting that a graph with with high number of nodes that are connected with nodes of low degree can have the same Neighbors Average Degree with a graph that has a low number of nodes connected to nodes with high degree. 
For this reason this metrics should be used in accordance with other graph indicators. 
In figure~\ref{fig:knn} we plot the boxplots of the nodes for each measured time period. 
The vertical lines correspond to the interquartile range and the dot to the median value. 
As with other structural measures, we notice a familiar pattern where there is an increase that peaks at mid 2009 and drops to approximately 50 with small fluctuations until the end of 2014.

\subsection{Motifs RAND-ESU}
\label{sec:motifs}
\begin{figure}[t]
  \centering
  \includegraphics[width=0.48\textwidth]{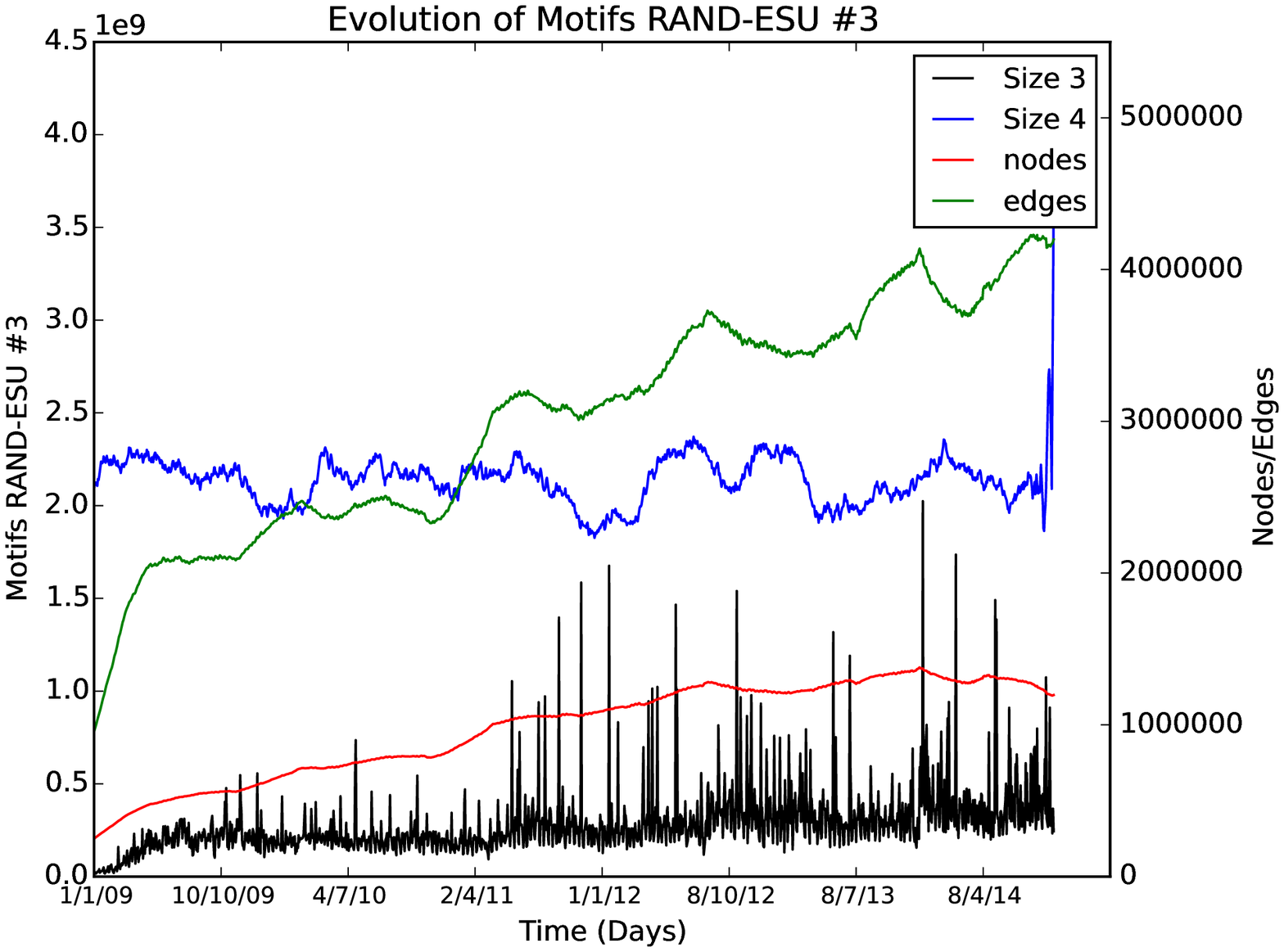}\llap{\makebox[12.0cm]{\raisebox{3.95cm}{\includegraphics[height=2cm]{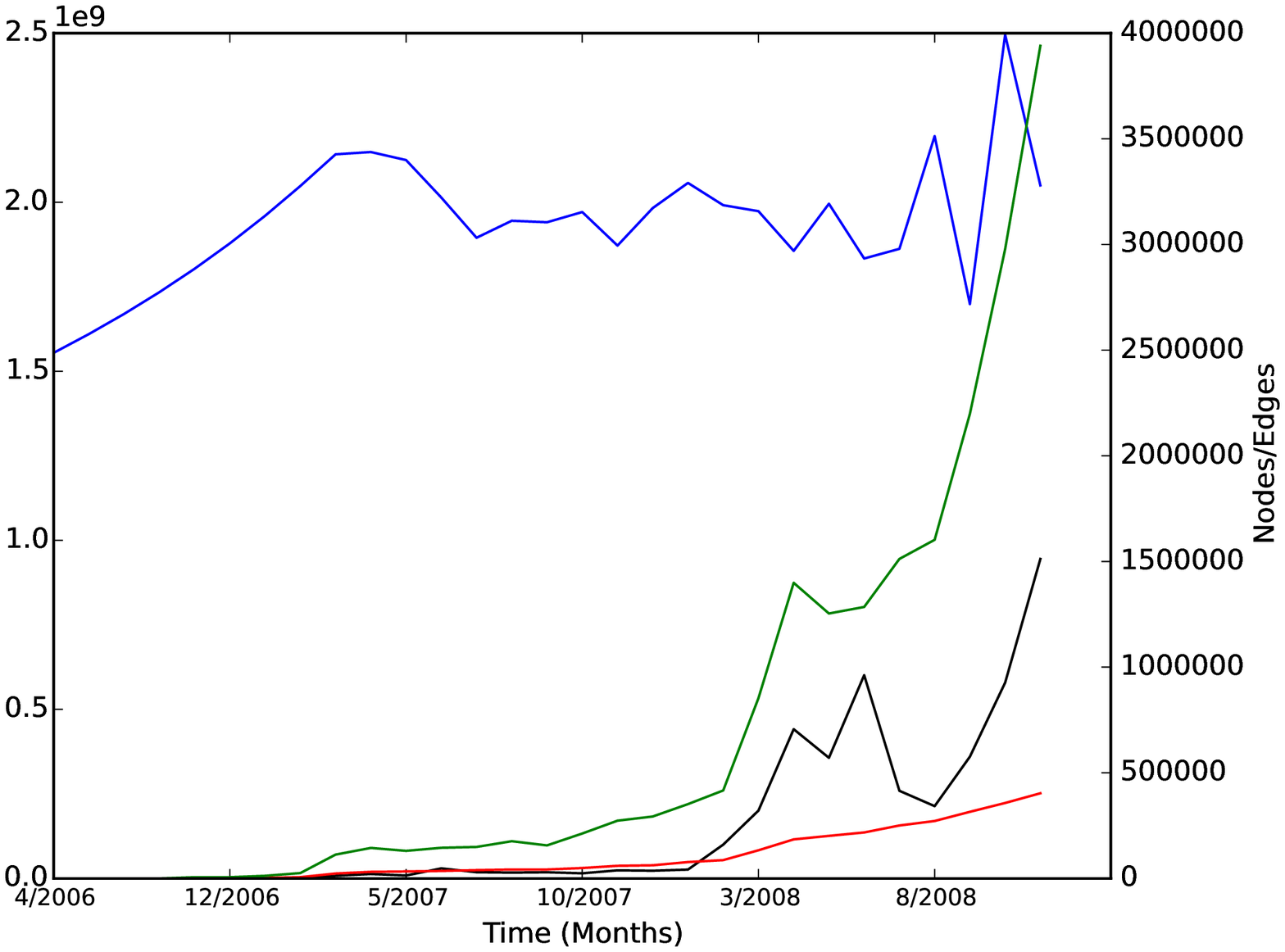}}}}
  \caption{Number of Motifs identified with the RAND-ESU method per day from start of 2009 to the end of 2014 and per month from April 2006 to December 2009 (embedded graph).}
  \label{fig:motifs}
\end{figure}

Motifs are small structured topologically equivalent subnetworks. 
Small motifs can play an important role on the functionality of networks and this has been demonstrated mainly on biological networks. 
It is an open question whether the presence of a small or large number of small motifs alters the growth dynamics, functionality or other characteristics of OSNs. 
Here, we apply the RAND-ESU method for locating small motifs of size 3 and 4~\cite{Wernicke2006}.
On figure~\ref{fig:motifs} we plot the number of motifs of size 3 (black line) and 4 (blue line).
The number of motifs is very large and can reach the number of billions.
Nevertheless we notice a steady increase of motifs with size 3 that peaks again on mid 2009 and a stabilization for the subsequent time periods with increasing fluctuations.
On the other hand motifs of size 4 show steady values around 2 billions. 
Moreover, the fluctuations as expected is higher than 3 sized motifs (on the plot we have a applied a smoothing parameter for visualization purposes).

\subsection{PageRank}
\label{sec:pagerank}
\begin{figure}[t]
  \centering
  \includegraphics[width=0.48\textwidth]{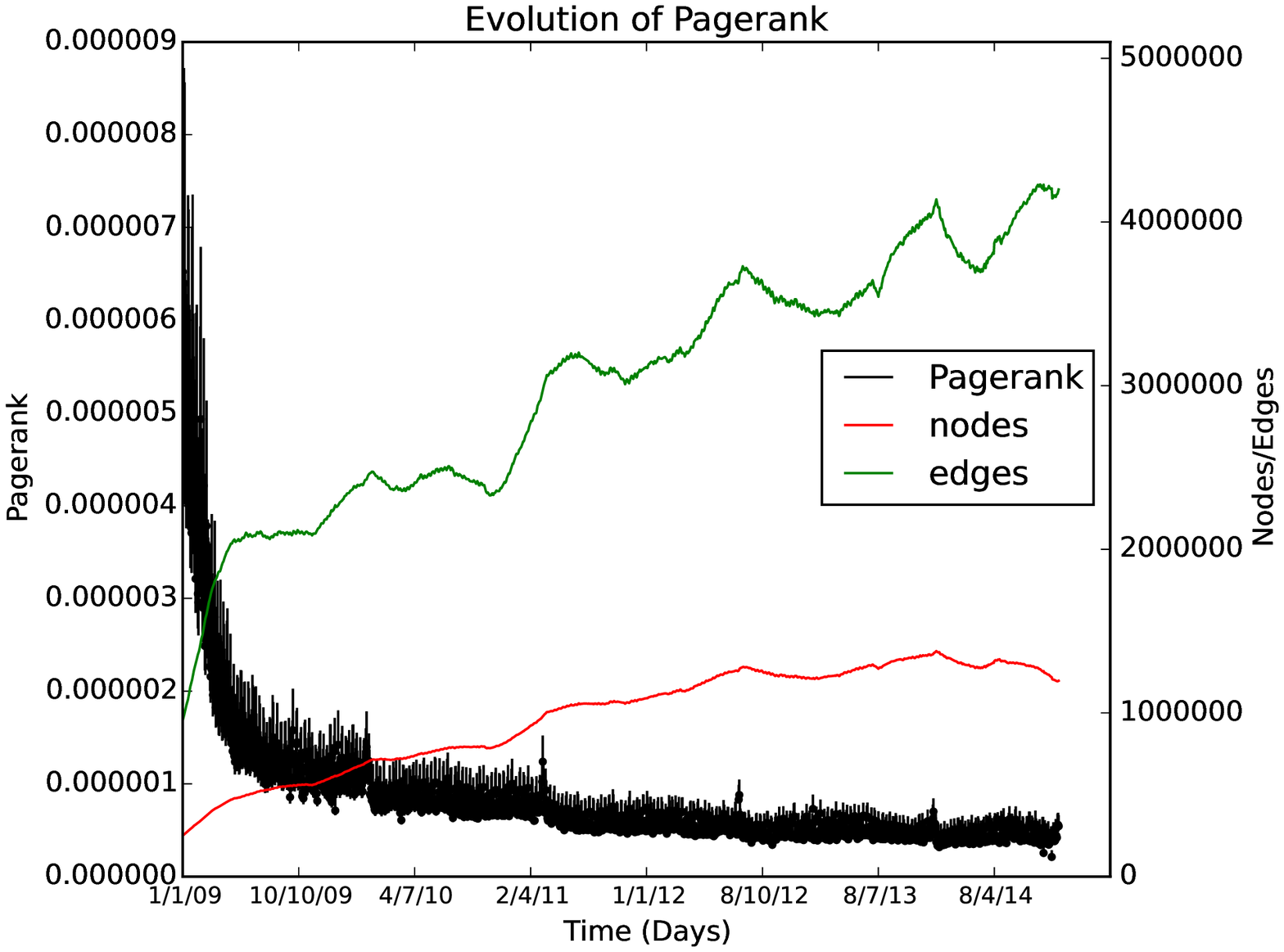}\llap{\makebox[11.5cm]{\raisebox{3.95cm}{\includegraphics[height=2cm]{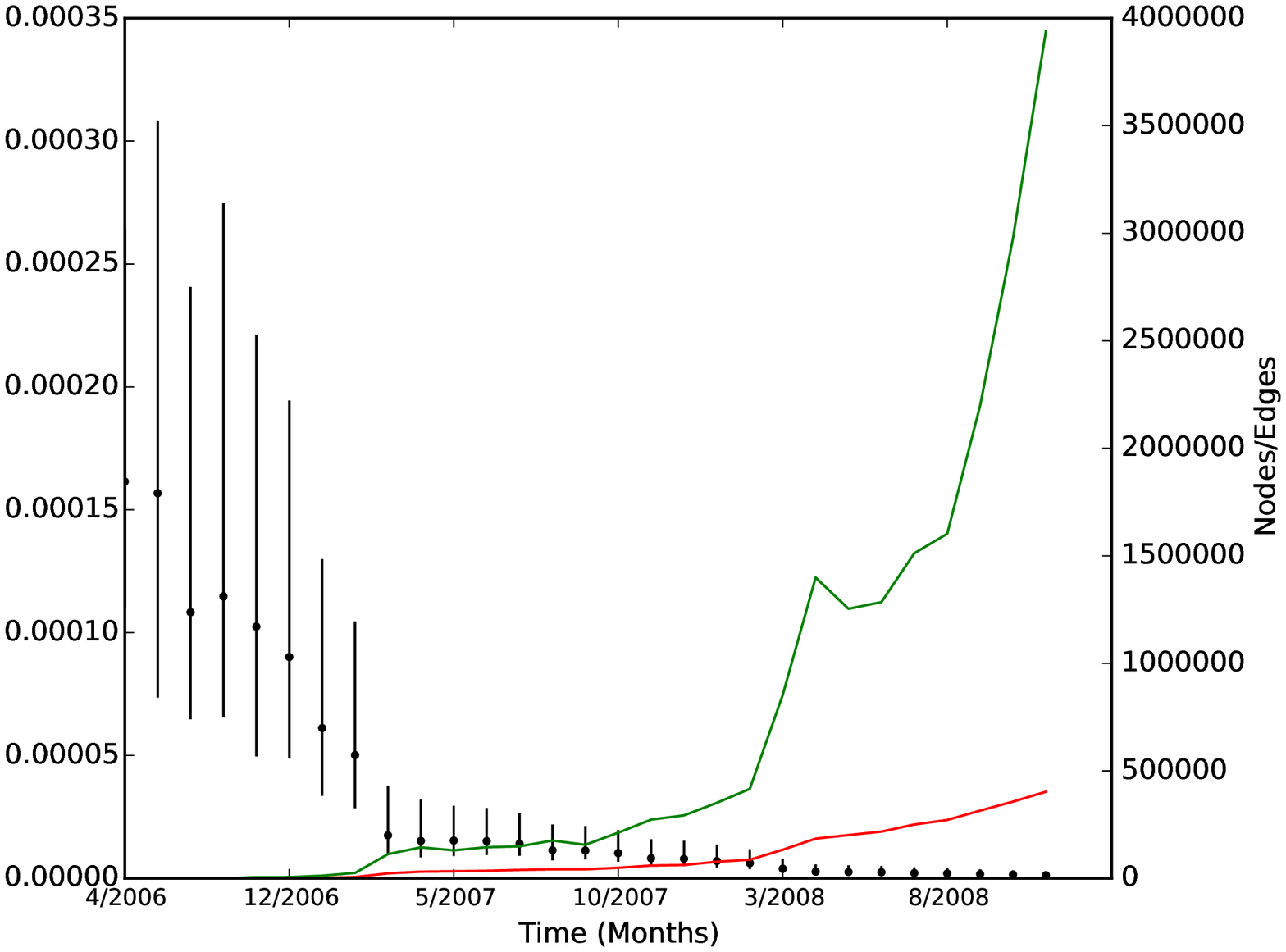}}}}
  \caption{Boxplots of Pagerank per day from start of 2009 to the end of 2014 and per month from April 2006 to December 2009 (embedded graph).}
  \label{fig:pagerank}
\end{figure}

Pagerank is one the most popular metric for measuring a user's (or a page's) influence mainly because it was introduced and adopted successfully by Google~\cite{Brin1998}.
There is an alternative of PageRank specially designed for Twitter users called TwitterRank~\cite{Weng2010} (http://tunkrank.com/) that takes into account retweets, mentions along with other metrics.
An extensive presentation on the application of this measure on Twitter can be found at \cite{10.1371/journal.pone.0084265}.
On this study the authors identified two fundamentally different types of Twitter users with different PageRank attributes.
Type 1 users have many followers but are not following many other users and Type 2 users that are also followed by many users but they also follow, in turn, many users.
The general principal of PageRank is that to each node we assign a value that is proportional to the sum of the PageRank value of the nodes that are connected to it.
Nodes without connecting nodes have a PageRank value of 1.
This procedure is recursively calculated for all nodes. 
On Figure~\ref{fig:pagerank} we plot the boxplot bars of the PageRank values throughout time. 
From this plot is evident that the median PageRank value of the graph is constantly dropping starting from the very early periods of Twitter.
Around mid 2009 the drop is becoming more stable and shows a convergence trend towards little above zero. 
We also notice that the variation of this measure is decreased.

\subsection{Similarity Inverse Log Weighted}
\label{sec:similarityinverselogweighted}
\begin{figure}[t]
  \centering
  \includegraphics[width=0.48\textwidth]{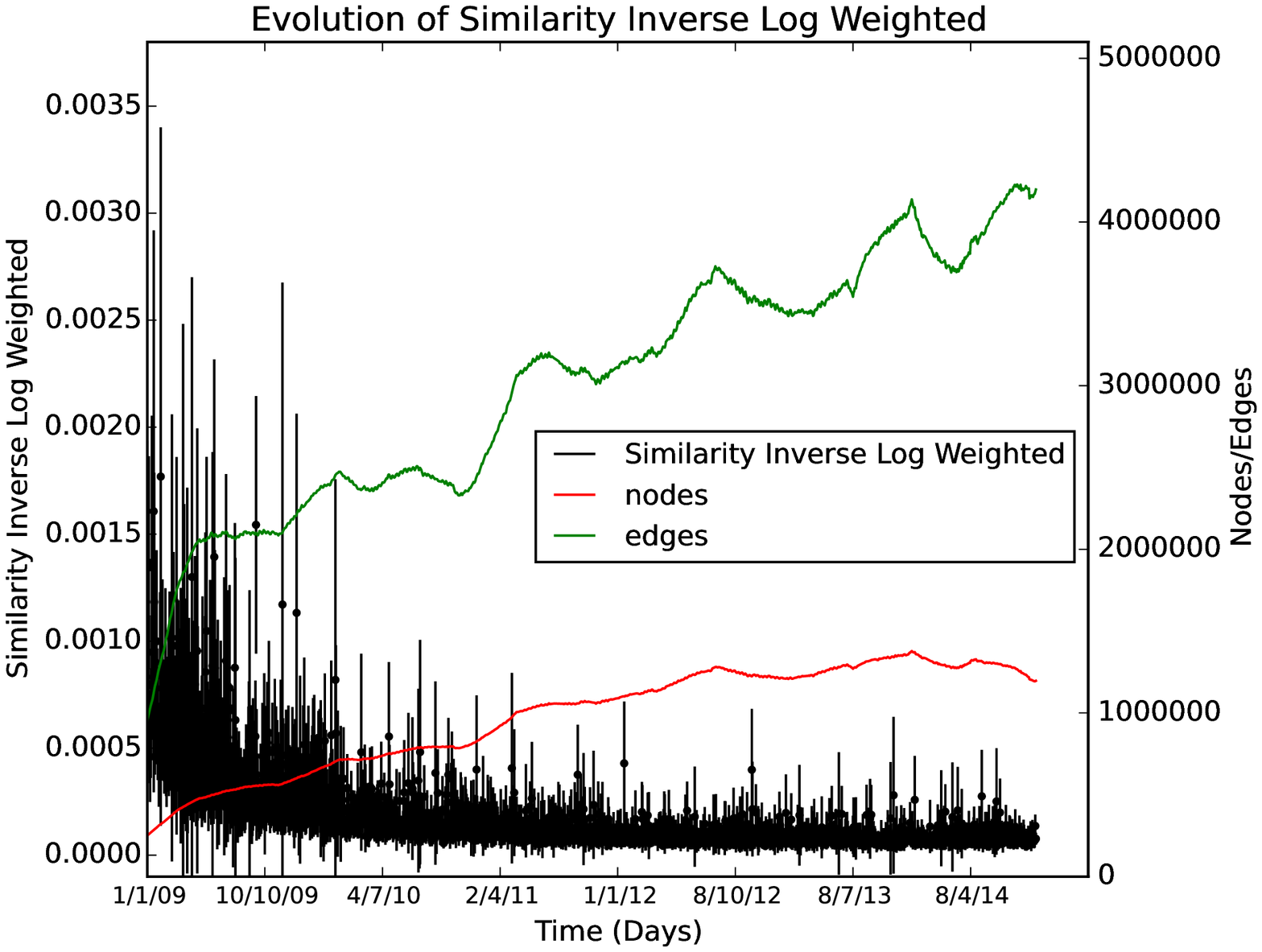}\llap{\makebox[11.5cm]{\raisebox{3.95cm}{\includegraphics[height=2cm]{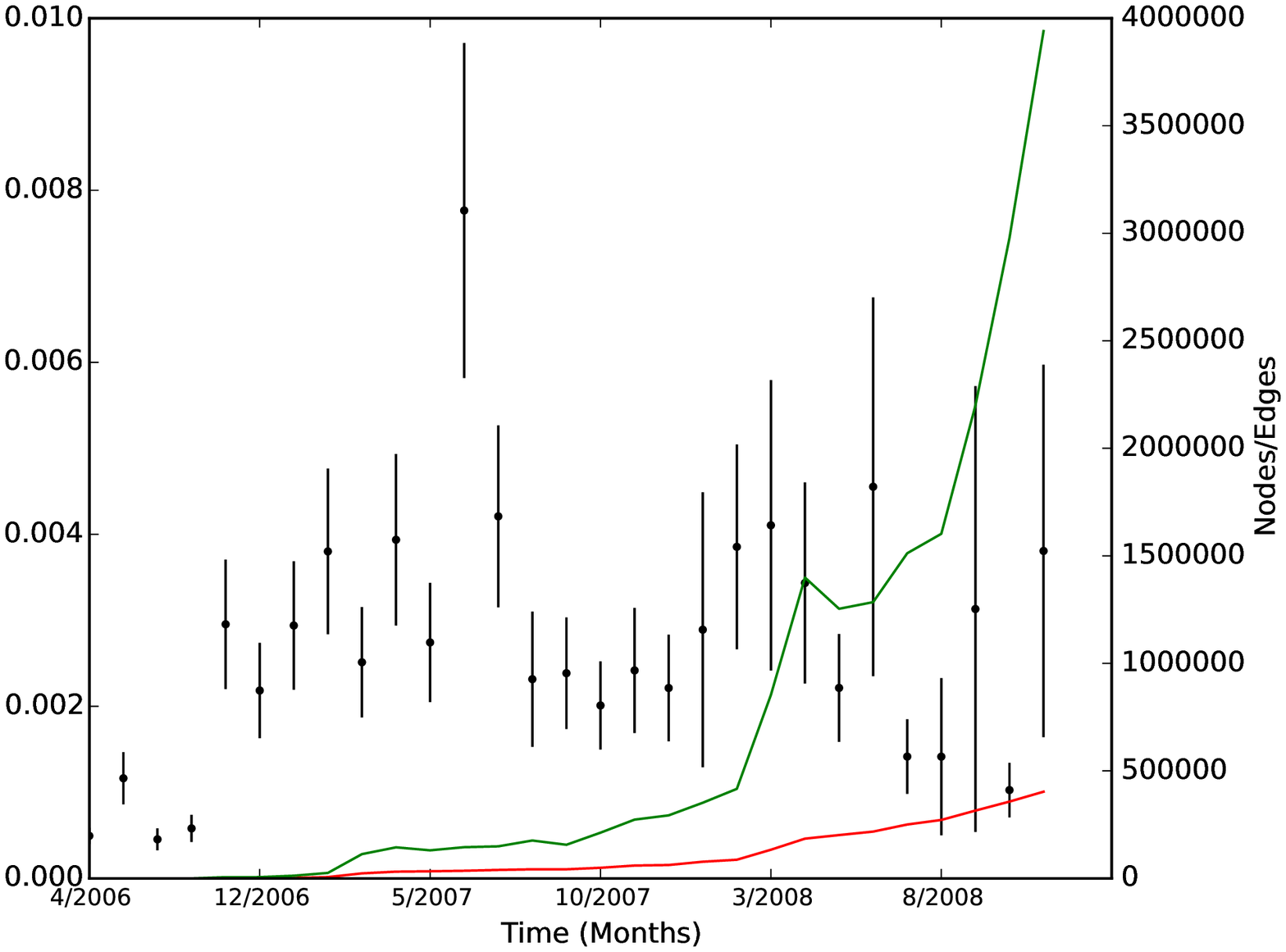}}}}
  \caption{Boxplots of average SILW per day from start of 2009 to the end of 2014 and per month from April 2006 to December 2009 (embedded graph).}
  \label{fig:similarityinverselogweighted}
\end{figure}

This metric (also referred as SILW) is defined as follows:
We assign a value to each node:
$$\frac{1}{\log(degree)}$$
Then for each pair on the graph we compute the sum of this value on their common neighbours.
This node similarity measure is based on the intuition that two nodes should be considered more similar if they share neighbours with low degrees.
Having common neighbours that are of high degree gives little or sometime no information on node similarity.
This metric returns a similarity list for all nodes, or else a list of lists.
For each node we measure the mean similarity to all other nodes.
The we plot the boxplot of these means of each time period.
On figure~\ref{fig:similarityinverselogweighted} we notice that on the early days of Twitter the nodes showed higher pairwise similarity.
This can be attribute to the fact that there were more tight sub-communities.
While time goes by the average pairwise similarity seems to decrease and converge on a slightly above zero value.
This is a sign of decrease of tight connected communities as the network grows.

\subsection{Transitivity}
\label{sec:transitivity}
\begin{figure}[t]
  \centering
  \includegraphics[width=0.48\textwidth]{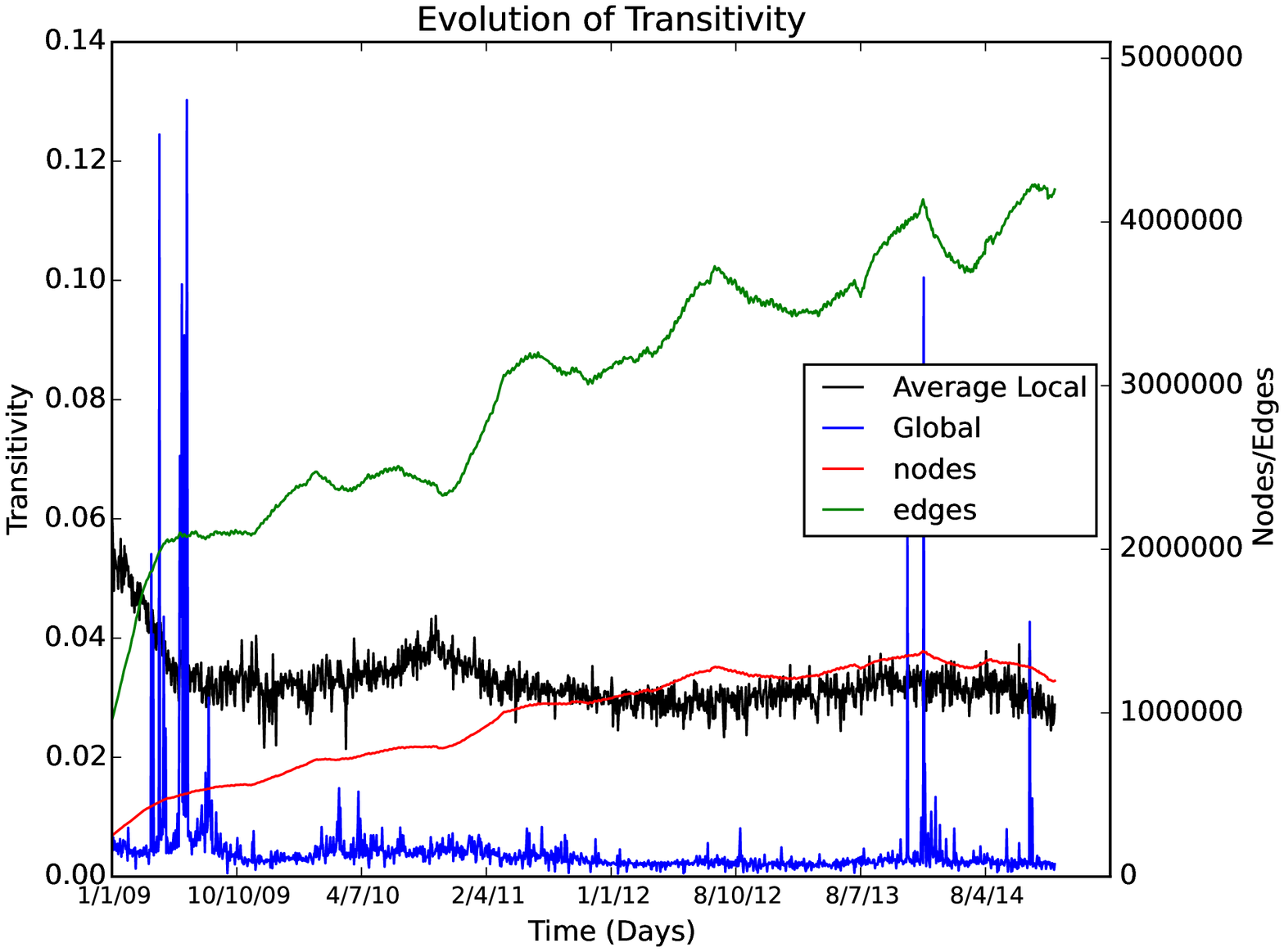}\llap{\makebox[11.5cm]{\raisebox{3.95cm}{\includegraphics[height=2cm]{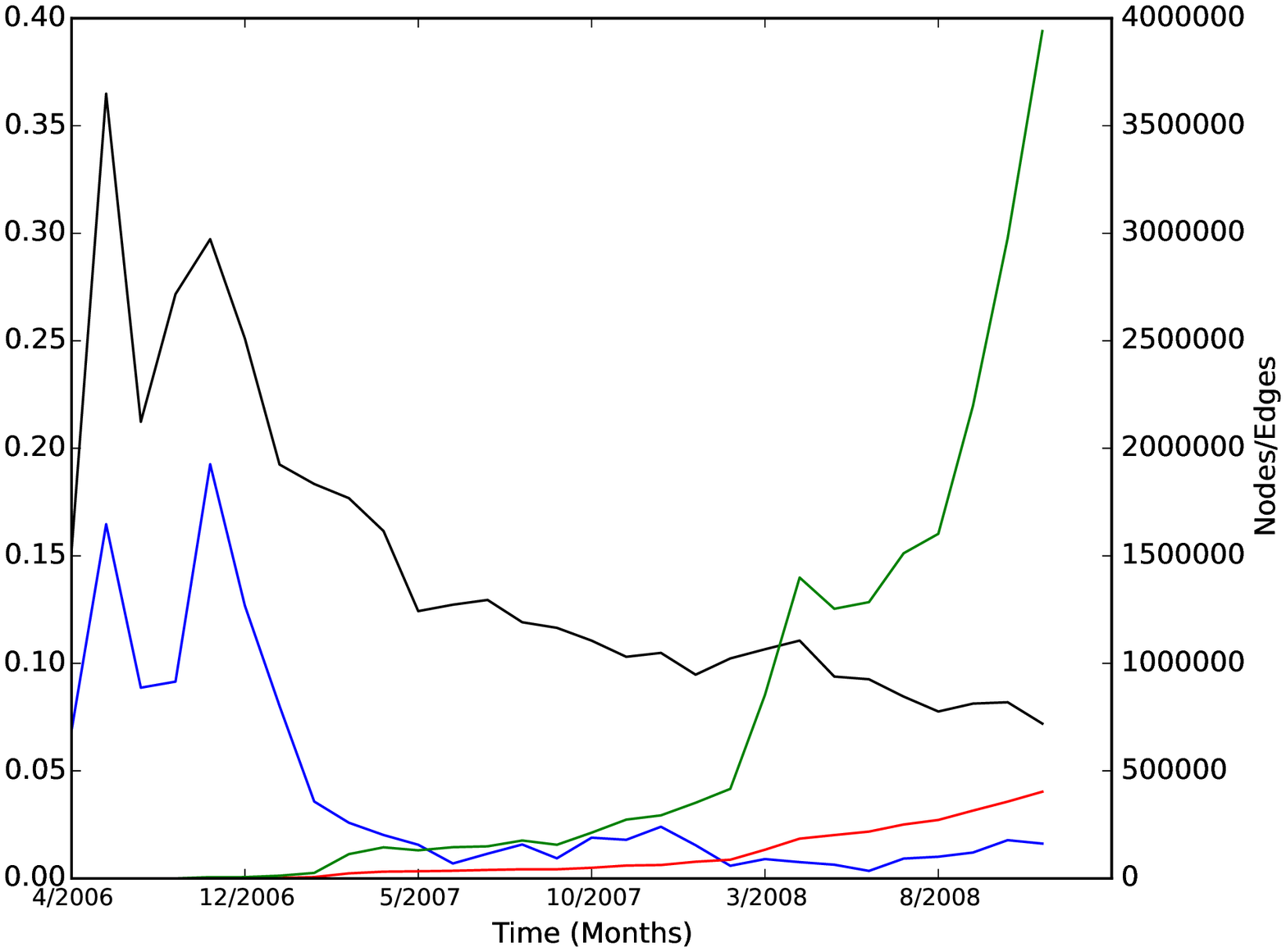}}}}
  \caption{Local and Global evaluations of Transitivity (Clustering Coefficient) per day from start of 2009 to the end of 2014 and per month from April 2006 to December 2009 (embedded graph).}
  \label{fig:transitivity}
\end{figure}

Transitivity (or else ``clustering coefficient") measures the connectivity of local communities.
It is the probability that two neighbours of a node are themselves connected. 
There are two flavors of this metric, the local and the global. 
The local transitivity of a node measures the ratio of edges within the neighborhood of this node, to the number of maximum possible edges in the same neighborhood. 
Here we report the average local transitivity over all nodes. 
The global transitivity measures the ratio of closed triplets to the number of connected triplets of nodes in the graph.
A closed triplet is when a node is part of a fully connected triangle. 
A connected triplet is when three nodes are connected by two or three edges. 
Local transitivity has been traditionally a measure of the `small-world' attribute of the graph.
Global transitivity is indicative of the clustering attribute of the graph. 
In our measurements both flavors assume that the graph is undirected and shown on figure~\ref{fig:transitivity}.
The average local transitivity shows a decline starting from the beginning of Twitter and stabilizes on values close to 0.03 at later periods. 
The global transitivity seems to be steady apart from some incidental spikes.  

\section{Discussion}
\label{sec:discussion}

\subsection{The Inflation and Deflation of Twitter}
\label{sec:inflaion}

The general conclusions from our measurements is that the structure of Twitter has undergone three major periods.
The first period is from the beginning (April 2006) until middle of 2009.
In this period we notice an explosive-like increase in all metrics that measure the dynamics of the network.
The betweenness centrality, the average degree, the Kleinberg's hub score, the KNN average degree and SILW all seem to have peaked at the end of this period. 
The impressive growth of these measures reflect the increase in popularity of Twitter both among existing users (which created more connections) and the attraction of new. 

The second period is from middle 2009 to start of 2011.
On this period Twitter shows a deflation of measures that are associated with growth dynamics. 
This can be attributed to a natural return to normality, or else correction, where Twitter peaked in popularity and started growing in more natural rates. 
External factors (like blocking of Twitter in China) might have also contributed to this~\cite{twittergrowth}.

The third period is from start of 2011 to at least the end of 2014.
During this time the growth follows stable rates and the compactness of the graph does not seem to change.
This trend also shows that this period of Twitter is going to last for a long period. 

Other metrics like the closeness, cocitation, density, eccentricity, eigenvector centrality, pagerank and local transitivity show a constant decrease tense (although not all with the same rate).
These metrics are associated more to the influence of individual nodes rather to the general structure of the graph.
This shows the transition of Twitter as a medium from a niche social network for microblogging, to a more generic medium for all kinds of online interaction. 

\subsection{Limitations and Future work}

Although evaluated, some metrics weren't able to produce any useful information. These metrics were the strength and the edge-connectivity.

The strength of a node is the sum of the weights of its edges.
Since our graph does not contain edge weights we do not include this metric. 
Nevertheless it is interesting to include weights values that reflect meta-information regarding an edge (for example retweets or mentions) and check the evolution of this metric.

The edge connectivity between two nodes measures the number of edges that have to be removed from the graph in order to disconnect them.
This measure is applied only in connected graph. 
Since our graph, due to the time splitting, contains small unconnected components, we weren't able to evaluate this metric.
As a future work we plan to preprocess our graph by extracting the largest graph component and apply this measure.

It is also essential to note that generic graph metrics like these that we study in this paper can sometimes be inadequate for the study of some aspects of OSNs. 
For example \cite{bringmann2010learning} studied the graph structure in a microscopic level mining for local patterns that play pivotal role in graph evolution. 
Another limitation of our methods is that our graph sampling technique (Random Nodes) might under represent weekly connected components. 
This sampling technique is used when a metric cannot be applied to the complete graph (this happens in 6 out of 24 metrics). 
Another limitation is that our sample size is inferior compared to other studies. 
This is mainly due to the harsh limitations of the Twitter's API. 
To remedy this, we have also collected the social graph of the SNAP dataset \cite{yang2011patterns} that contains 40 million users and we plan to apply on it these measurements. 
We have also collected the meta-information of 250 million users (also called user objects) and we plan to investigate the correlation between the information available there (i.e. geographic location~\cite{kulshrestha2012geographic}) and the presented metrics.

\subsection{Final Remarks}
Graph metrics is an essential part in the field of social networks and graph theory in general.
In this paper we have demonstrated that there is a big variety of metrics that are sparsely used in social network studies and can be of extreme importance.
We also argue that a complete dimension of Twitter's OSN is under represented in these studies due to its unavailable from Twitter's API.
This is the time creation of the edges. 
We demonstrate how a simple (and already published) heuristic can approximate this creation time thus contributing to time analysis of Twitter's OSN.
Finally we argue that although the computational nature of some of these metrics is prohibitive for even medium sized OSNs, a simple random sub-sampling can produce fair approximations of these values and enhance our knowledge on graph structure and evolution.

%% file: acknowledgements.tex
\section*{Acknowledgements}
\label{sec:Acknowledgements}

We would like to thank Marian Boguna and Kolja Kleineberg on the discussions and contribution on the infrastructure on the University of Barcelona. 
Also we would like to thank Hariton Efstathiades and Demetris Antoniades for their valuable comments.
This work was supported by the FP7 Marie-Curie ITN iSocial funded by the EC under grant agreement no 316808. This work was also supported by the: 
NSF Grant CNS-13-18415, FP7-PEOPLE-2010-IOF project XHUNTER,
 No. 273765,
 Prevention of and Fight against Crime Programme of the European Commission – Directorate-General Home Affairs (project GCC),
 European Union's Prevention of and Fight against Crime Programme “Illegal Use of Internet” e ISEC 2010 Action Grants, grant HOME/2010/ISEC/AG/INT-002.